\documentclass[]{aa}
\usepackage{graphicx,natbib}

\def\mV{\mbox{$M_{V}$}}

\def\aV{\mbox{$\rm A_V$}}

\def\jh{\mbox{$(J-H)$}}
\def\hk{\mbox{$(H-K_s)$}}
\def\jk{\mbox{$(J-K_s)$}}
\def\mMJ{\mbox{$(m-M)_J$}}
\def\mMo{\mbox{$(m-M)_O$}}
\def\ebv{\mbox{$E(B-V)$}}

\def\ejh{\mbox{$E(J-H)$}}

\def\rc{\mbox{$R_{\rm c}$}}
\def\rch{\mbox{$R_{\rm c,H}$}}
\def\rl{\mbox{$R_{\rm RDP}$}}
\def\rx{\mbox{$R_{\rm ext}$}}

\def\rth{\mbox{$R_{\rm t,H}$}}

\def\ms{\mbox{$M_\odot$}}
\def\ds{\mbox{$d_\odot$}}
\def\rs{\mbox{$R_\odot$}}
\def\dgc{\mbox{$R_{\rm GC}$}}

\def\dSC{\mbox{$\Delta R_{\rm SC}$}}

\def\jj{\mbox{$J$}}
\def\hh{\mbox{$H$}}
\def\ks{\mbox{$K_s$}}

\def\mcl{\mbox{$M_{\rm clu}$}}
\def\kms{\mbox{$\rm km\,s^{-1}$}}

\begin{document}

\title{The fate of the pre-main sequence-rich clusters Collinder\,197 and vdB\,92: 
dissolution?}

\author{C. Bonatto\inst{1} \and E. Bica\inst{1}}

\offprints{C. Bonatto}

\institute{Universidade Federal do Rio Grande do Sul, Departamento de Astronomia\\
CP\,15051, RS, Porto Alegre 91501-970, Brazil\\
\email{charles@if.ufrgs.br, bica@if.ufrgs.br}
\mail{charles@if.ufrgs.br} }

\date{Received --; accepted --}

\abstract
{Since most of the star clusters do not survive the embedded phase, their early  
dissolution appears to be a major source of field stars. However, catching a
cluster in the act of dissolving is somewhat elusive.}
{We investigate the nature and possible evolution of the young Galactic star 
clusters Collinder\,197 (Cr\,197) and vdB\,92.}
{Photometric and structural properties are derived with near-infrared photometry 
and field-star decontamination. Kinematical properties are inferred from proper 
motions of the main sequence (MS) and pre-MS (PMS) member stars.}
{The colour-magnitude diagrams (CMDs) are basically characterised by a poorly-populated 
MS and a dominant fraction ($\ga75\%$) of PMS stars, and the combined MS and PMS CMD 
morphology in both clusters consistently constrains the age to within $5\pm4$\,Myr, 
with a $\sim10$\,Myr spread in the star formation process. The MS$+$PMS stellar masses 
are $\approx660^{+102}_{-59}\,\ms$  (Cr\,197) and $\approx750^{+101}_{-51}\,\ms$ (vdB\,92). 
Cr\,197 and vdB\,92 appear to be abnormally large, when compared to clusters within the 
same age range. They have irregular stellar radial density distributions (RDPs) with a 
marked excess in the innermost region, a feature that, at less than 10\,Myr, is more 
likely related to the star formation and/or molecular cloud fragmentation than to 
age-dependent dynamical effects. The velocity dispersion of both clusters, derived from 
proper motions, is in the range $\sim15 - 22\,\kms$.}
{Both clusters appear to be in a super-virial state, with velocity dispersions higher 
than those expected of nearly-virialised clusters of similar mass and size. A possible 
interpretation is that Cr\,197 and vdB\,92 deviate critically from dynamical equilibrium, 
and may dissolve into the field. We also conclude that early cluster dissolution leaves 
detectable imprints on RDPs of clusters as massive as several $10^2\,\ms$. Cr\,197 and 
vdB\,92 may be the link between embedded clusters and young stellar associations. }

\keywords{{\em (Galaxy:)} open clusters and associations: general; {\em (Galaxy:)} 
open clusters and associations: individual: Collinder\,197 and vdB\,92}

\titlerunning{Collinder\,197 and vdB\,92: Dissolving star clusters?}

\maketitle

\section{Introduction}
\label{Intro}

It is now well-established that a significant fraction of the embedded star clusters 
dissolve into the field on a time-scale of a few $10^7$\,Myrs. Basically, dissolution 
occurs mainly because the gravitational potential can be rapidly reduced by internal 
processes, such as the impulsive gas removal by supernovae and massive star winds 
associated with this early period. As a consequence, an important fraction of the stars, 
especially of low mass, end up moving faster than the scaled-down escape velocity, and 
may be lost to the field (e.g. \citealt{GoBa06}). This process can dissolve the very young 
star clusters on time-scale of $10-40$\,Myr. 

From the stellar content perspective, the early cluster dissolution depends essentially 
on the effective star-formation efficiency, the total mass converted into stars, 
and the mass of 
the more massive stars (e.g. \citealt{tutu78}; \citealt{GoBa06}). However, there is
also evidence indicating that the determining factor for cluster survival during gas 
expulsion is the virial state of the stars just before the onset of this phase, so 
that clusters formed with a dynamically cold stellar component are more likely to 
survive (\citealt{GoodW09}).

In any case, the early dissolution of embedded clusters may lead to the formation 
of OB stellar groups (e.g. \citealt{Goul00}), the subsequent dispersion of which may 
be an important source of field stars (e.g. \citealt{Massey95}). Indeed, \citet{LL2003} 
suggest that only about 5\% of the Galactic embedded clusters dynamically evolve into 
gravitationally bound open clusters (OCs). With such a high dissolution rate, the 
embedded clusters could be the major contributors of field stars in galaxies for 
generations. However, recent studies suggest that the early dissolution rate in the 
Magellanic Clouds is significantly lower ($\la30\%$ - \citealt{deGG08}; \citealt{deGG09}) 
than in the Milky Way (\citealt{LL2003}), or in other galaxies such as the Antennae
(e.g. \citealt{Whit07}) or M\,51 (e.g. \citealt{Bast05}).

Observationally, low-mass star clusters younger than $\sim10$\,Myr, in general, have
Colour-Magnitude Diagrams (CMDs) with an under-populated and developing main sequence 
(MS) and a more conspicuous population of pre-MS (PMS) stars. Typical examples are 
NGC\,6611, NGC\,4755, NGC\,2239, NGC\,2244, Bochum\,1, Pismis\,5, NGC\,1931, vdB\,80, 
and BDBS\,96 (\citealt{Pi5} and references therein). In terms of CMD morphology, bound 
and un-bound young clusters are expected to present similar evolutionary sequences. On 
the other hand, the important early changes in the potential that affect the large-scale 
internal structure of clusters should be reflected on the stellar radial density profile 
(RDP). Bochum\,1 (\citealt{Bochum1}) and NGC\,2244 (\citealt{N2244}), for instance, appear 
to be representatives of this scenario (i.e. structures evolving towards dissolution in a 
few $10^7$\,yr), in which an irregular RDP cannot be represented by a cluster-like (i.e. 
an approximately isothermal sphere) profile. In this context, irregular RDPs in young
clusters - when coupled to an abnormally high velocity dispersion - may reflect significant 
profile erosion or dispersion of stars, and point to important deviations from dynamical 
equilibrium.  

The rather complex interplay among environment conditions, effective star-formation 
efficiency (as defined in \citealt{GoBa06}), 
and total mass converted into stars, is probably what explains the difference between 
(dissolving) objects like Bochum\,1 and NGC\,2244, and bound young OCs (in which the 
MS$+$PMS stars distribute according to a cluster RDP as in NGC\,6611 and NGC\,4755). 
Consistent with the above mass-dependent scenario, the MS$+$PMS mass of Bochum\,1 and 
NGC\,2244 is a factor 2-3 lower than in NGC\,6611 and NGC\,4755. 

In this paper we investigate the nature of the poorly-studied, young (age $\sim5$\,Myr; 
Sect.~\ref{DFP}), large (with radii within $8-12$\,pc; Sect.~\ref{struc}), and PMS rich 
(with stellar masses within $660 - 750\,\ms$; Sect.~\ref{MF}) clusters Cr\,197 and vdB\,92. 
Our main goal is to determine whether such young stellar 
systems can be characterised as typical OCs or if they are heading towards dissolution. 
In addition, we will derive their fundamental and structural parameters, most of these 
for the first time. 

\begin{figure}
\begin{minipage}[b]{0.50\linewidth}
\includegraphics[width=\textwidth]{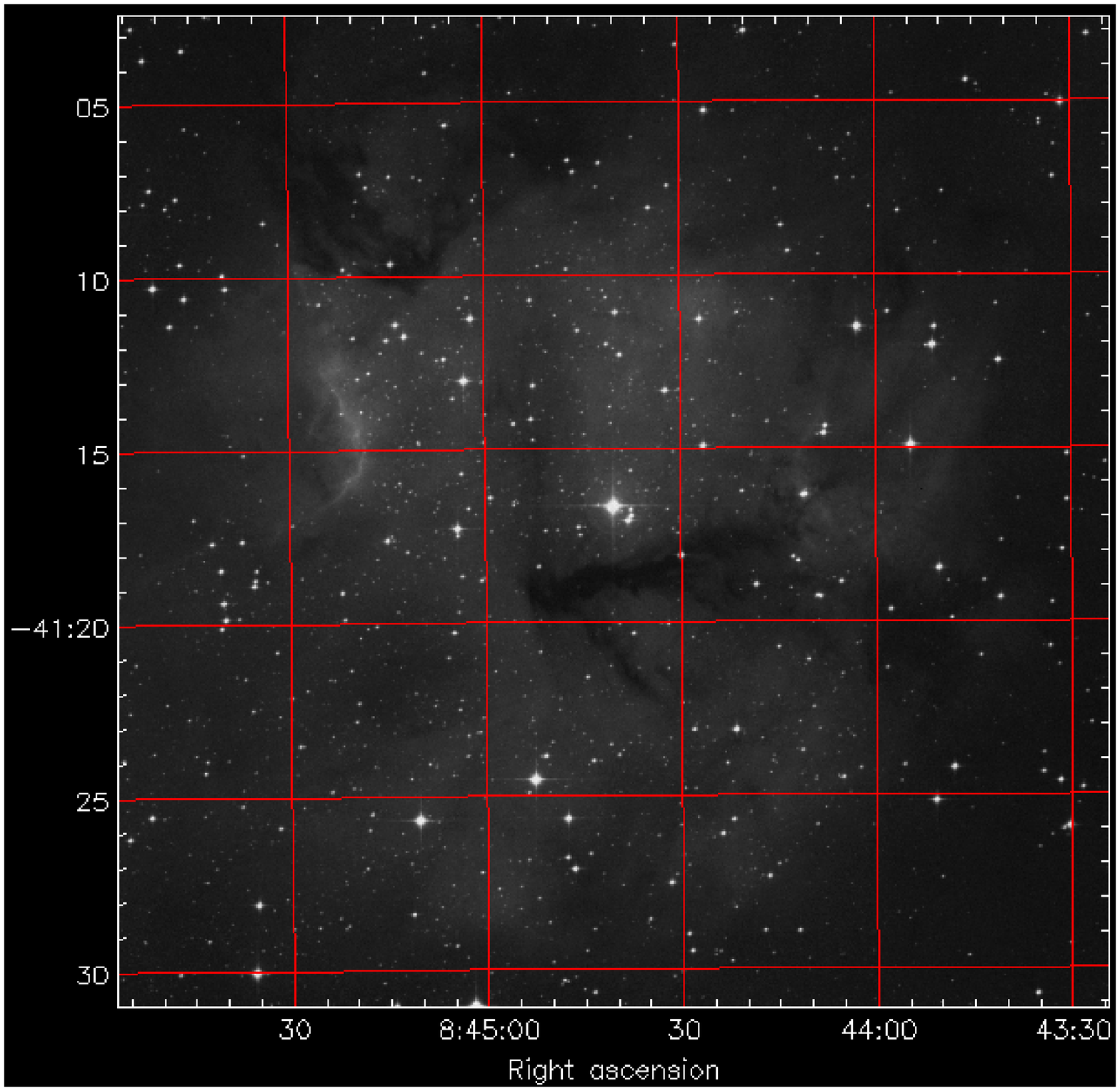}
\end{minipage}\hfill
\begin{minipage}[b]{0.50\linewidth}
\includegraphics[width=\textwidth]{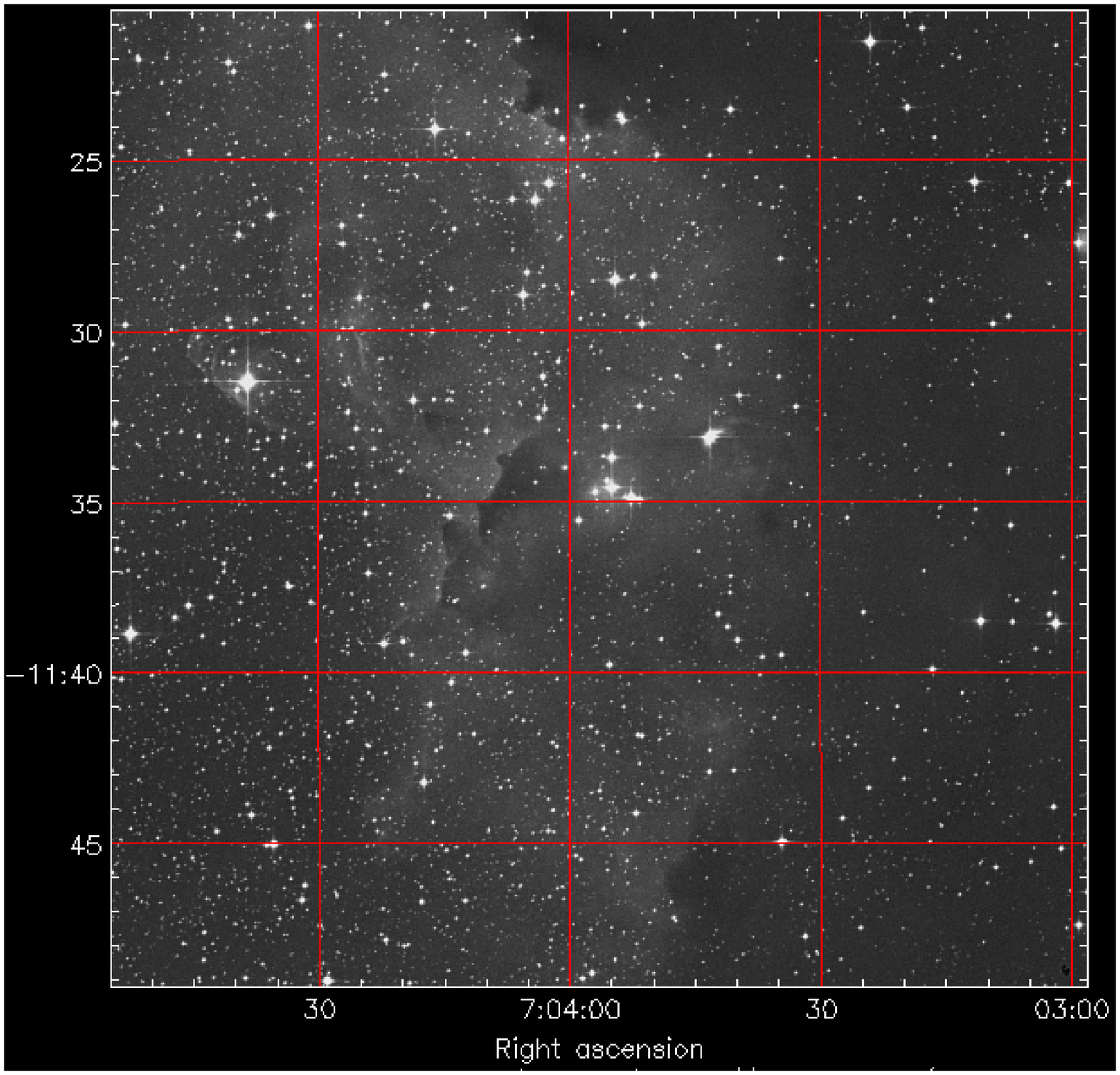}
\end{minipage}\hfill
\begin{minipage}[b]{0.50\linewidth}
\includegraphics[width=\textwidth]{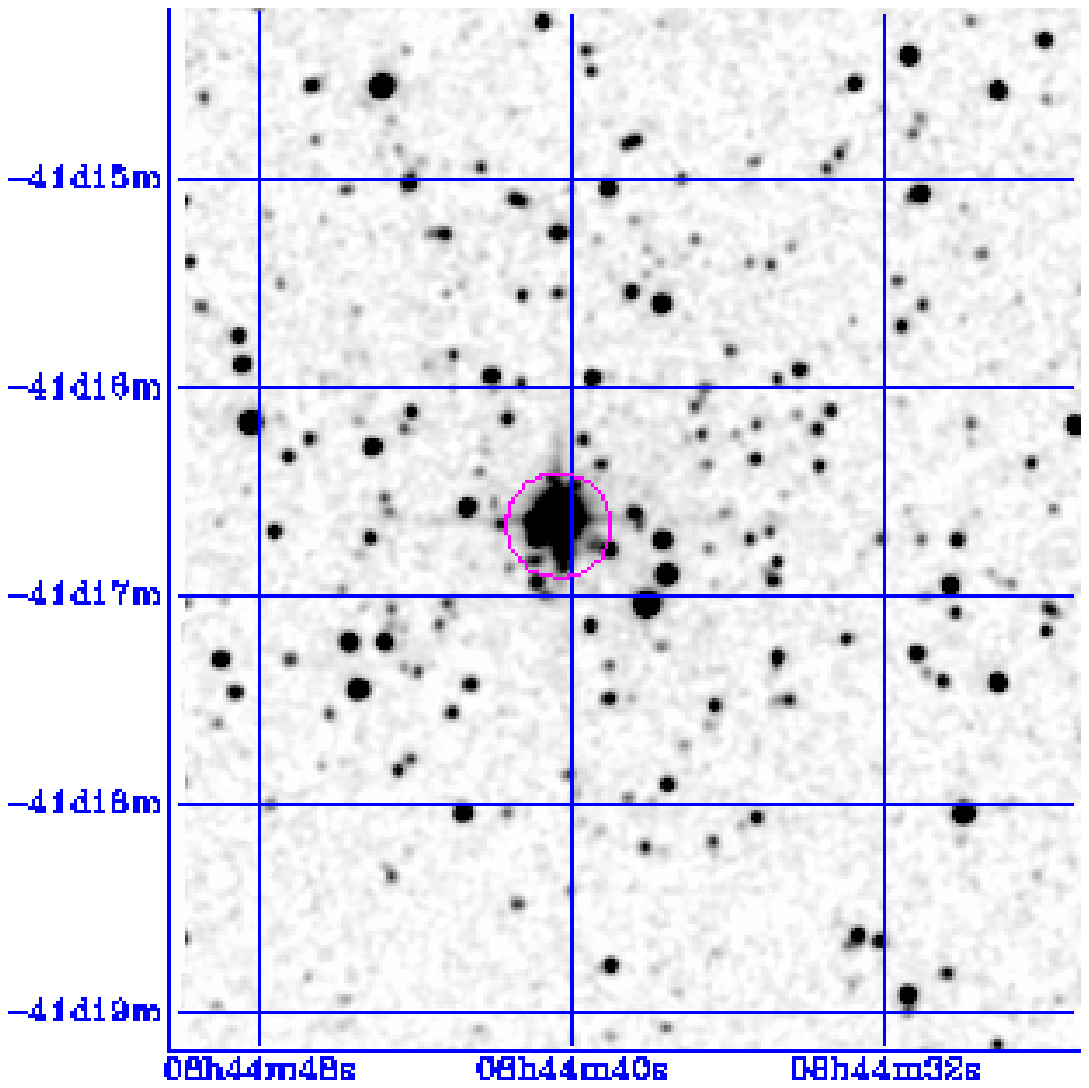}
\end{minipage}\hfill
\begin{minipage}[b]{0.50\linewidth}
\includegraphics[width=\textwidth]{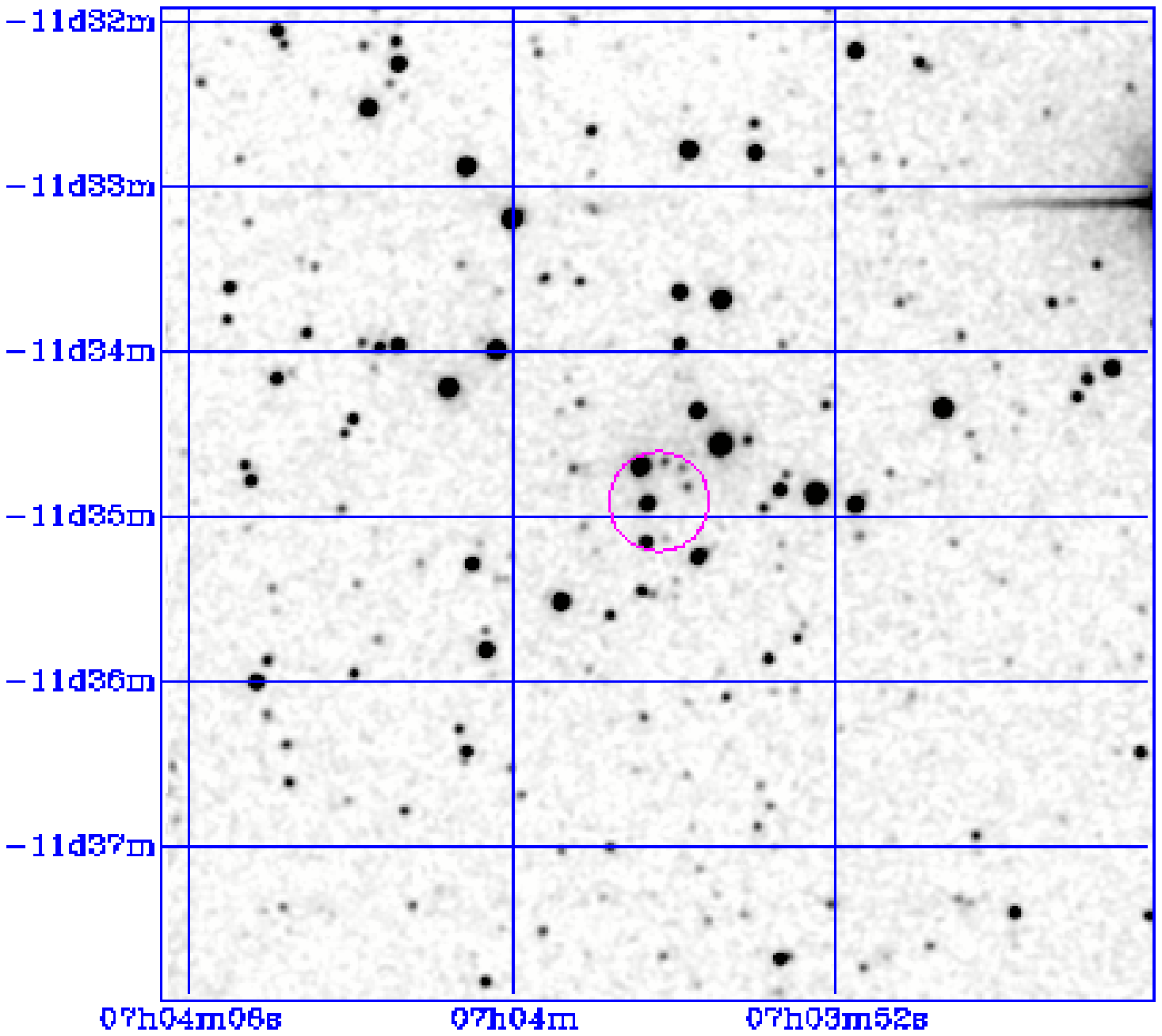}
\end{minipage}\hfill
\caption[]{Top: $20\arcmin\times20\arcmin$ DSS-II R images centred on Cr\,197 (left) 
and vdB\,92 (right). Gas emission, dust reflection and/or absorption are present 
in the fields in varying proportions. Bottom: 2MASS \ks\ images covering 
$5\arcmin\times5\arcmin$ (Cr\,197) and $6\arcmin\times6\arcmin$ (vdB\,92).  
Orientation: North to the top and East to the left.}
\label{fig1}
\end{figure}

\begin{table*}
\caption[]{Fundamental parameters}
\label{tab1}
\tiny
\renewcommand{\tabcolsep}{0.73mm}
\renewcommand{\arraystretch}{1.25}
\begin{tabular}{cccccccccccccccc}
\hline\hline
&\multicolumn{6}{c}{WEBDA}&&\multicolumn{8}{c}{This work}\\
\cline{2-7}\cline{9-16}
Cluster&$\alpha(2000)$&$\delta(2000)$&Age&\ebv&\ds&D&&$\alpha(2000)$&$\delta(2000)$&
        $\ell$&$b$&Age&\ebv&\ds&\dSC\\
 & (hms)&($\degr\,\arcmin\,\arcsec$)&(Myr)&(mag)&(kpc)&(\arcmin)&&(hms)&($\degr\,\arcmin\,\arcsec$)
 &(\degr)&(\degr)&(Myr)&(mag)&(kpc)&(kpc)\\
(1)&(2)&(3)&(4)&(5)&(6)&(7)&&(8)&(9)&(10)&(11)&(12)&(13)&(14)&(15)\\
\hline
Cr\,197&08:44:51&$-$41:14:00&13&0.55&0.84&17.0&&08:44:40.3&$-$41:16:48.4&261.51&$+$0.94
         &$5\pm4$&$0.34\pm0.16$&$1.05\pm0.20$&$0.23\pm0.04$\\
vdB\,92&07:03:54&$-$11:29:23&40&0.25&1.50&24.0&&07:03:56.4&$-$11:34:54.7&224.66&$-$2.52
       &$5\pm4$&$0.22\pm0.13$&$1.38\pm0.26$&$1.04\pm0.19$ \\
\hline
\end{tabular}
\begin{list}{Table Notes.}
\item Col.~7: Optical diameter; Col.~14: distance from the Sun; col.~15: distance from 
the Solar circle. WEBDA data for Cr\,197 is based on photometry from \citet{VM73}.
\end{list}
\end{table*}

This paper is organised as follows. In Sect.~\ref{RecAdd} we recall literature data 
on both objects. In Sect.~\ref{2mass} we discuss the 2MASS photometry and build 
the field-star decontaminated CMDs. In Sect.~\ref{DFP} we derive fundamental cluster 
parameters. In Sect.~\ref{struc} we derive structural parameters. In Sect.~\ref{MF} 
we estimate cluster mass. In Sect.~\ref{propMot} we build the intrinsic proper motion 
distribution. In Sect.~\ref{Discus} we compare  structural parameters and dynamical 
state with those of a sample of template OCs. Concluding remarks are given in 
Sect.~\ref{Conclu}.

\section{Previous data on Cr\,197 and vdB\,92}
\label{RecAdd}

Collinder\,197 (Cr\,197) was discovered by \citet{Coll31} in Vela. It was also listed 
in the OC catalogue of \citet{Alter70} as OCl-742, and the ESO/Uppsala Blue 
Plate Southern (\citealt{Lauberts82}) as ESO\,313\,SC13. 

\citet{VM73} centred the cluster on the B4\,II star HIP\,42908 at 
$\alpha(2000)=08^h44^m40.3^s$, $\delta(2000)=-41\degr16\arcmin38\arcsec$. This star,
which is reported as a double system in SIMBAD, is the brightest in the area with 
$V=7.32$. HIP\,42908 may be the  cluster centre in the infrared, since strong absorption 
is suggested by the B plate image seen to the south/west of HIP\,42908. \citet{VM73}
found the reddening $\ebv=0.58$ and the distance from the Sun $\ds=1.05$\,kpc.

Besides dust absorption and emission in the area, Cr\,197 is embedded also in the HII 
region Gum\,15 (\citealt{Gum55}), also known as RCW\,32 (\citealt{RCW60}). Assuming 
that\,HIP 42908 is a cluster member, it might responsible for the emission. Indeed,
\citet{PetBo94} have identified low-luminosity emission-line stars in Cr\,197 and in
the R-association Vela R2, which may be part of their low-mass population. They 
derive $\ds=1.1$\,kpc and estimate a few $10^6$\,yr of age. \citet{Bat91} and 
\citet{Bat94} estimated $6.3$\,Myr of age, $\ds=1$\,kpc, a total
mass of $\approx100\,\ms$, and the integrated absolute magnitude $\mV=-5.16$. 

Found by \citet{vdB66}, the embedded cluster vdB\,92 is related to the reflection 
nebula vdB-RN\,92, which is located in Canis Major. The cluster is also listed as 
FSR\,1188 (\citealt{Froeb07}) who, based on the \hh\ photometry, derived the core 
and tidal radii $\rch=1.1\arcmin$ and $\rth=24\arcmin$, and the total number of 
members $N=297$\,stars. \citet{Magakian03} also reports the reflection nebula and 
related cluster. It is immersed in the large angular size association CMa\,OB1 
(\citealt{Claria74a}; \citealt{Claria74b}), and in the reflection nebula association 
CMa\,R1 (\citealt{vdB66}). The equally extended emission nebula IC\,2177 or Gum\,2 
(\citealt{Gum55}) seems to permeate the complex. The relatively bright star at 
$\approx3.4\arcmin$ to the northeast of the cluster vdB\,92 is the Herbig Be star 
Z\,CMa, which has a bipolar outflow (\citealt{Poetzel89}). Undoubtedly, vdB\,92 is 
part of a star-forming complex, and  deserves further analyses.

With an approach that differs in various aspects from the present one,
\citet{Soares03} studied the brighter sequences of vdB\,92, deriving
an age of 5-7\,Myr, a mean visible absorption $\aV=4.4$, and $\ds\approx1.5$\,kpc.

A few bright stars mixed with nebular gas and/or dust emission are seen in the
$20\arcmin\times20\arcmin$ R images (Fig.~\ref{fig1}, taken from LEDAS\footnote{Leicester 
Database and Archive Service (LEDAS) DSS/DSS-II service on ALBION; 
{\em http://ledas-www.star.le.ac.uk/DSSimage}.}). Close-ups of the clusters are 
shown in the smaller field 2MASS \ks\ images. Table~\ref{tab1} provides parameters found 
in the literature and as derived here. The central coordinates were re-computed to match 
the absolute maxima present in the stellar surface densities (Sect.~\ref{DecOut}). We note
that the re-computed central coordinates of Cr\,197 are almost coincident with HIP\,42908.

\section{Construction of decontaminated CMDs}
\label{2mass}

Since both objects still retain part of the primordial gas and dust (Fig.~\ref{fig1}),
the near-IR provides the adequate depth to study them, especially the faint stellar
sequences. For this purpose, we work with the 2MASS\footnote{The Two Micron All Sky 
Survey, All Sky data release (\citealt{2mass1997}) - {\em http://www.ipac.caltech.edu/2mass/releases/allsky/}} 
\jj, \hh, and \ks\ photometry, which provides the spatial and photometric uniformity 
required for wide extractions that, in turn, result in high star-count statistics.

Our group has been developing analytical tools for the 2MASS photometry that allow
us to statistically disentangle cluster evolutionary sequences from field stars in 
CMDs. Decontaminated CMDs have been used to investigate the nature of star cluster 
candidates and derive their astrophysical parameters. In summary, we apply {\em (i)} 
field-star decontamination to uncover the intrinsic CMD morphology, essential for a 
proper derivation of reddening, age, and distance from the Sun, and {\em (ii)} 
colour-magnitude filters, which are required for intrinsic stellar RDPs, as well as 
luminosity and mass functions (MFs). In particular, the use of field-star decontamination 
in the construction of CMDs has proved to constrain age and distance more than the raw 
(observed) photometry, especially for low-latitude and/or bulge-projected OCs 
(e.g. \citealt{ProbFSR}, and references therein).

Photometry for both clusters was extracted from VizieR\footnote{\em
http://vizier.u-strasbg.fr/viz-bin/VizieR?-source=II/246} in a wide circular field of 
radius $\rx=80\arcmin$, which is adequate to allow determination of the background level 
(Sect.~\ref{struc}) 
and to statistically characterise the colour and magnitude distribution of the field 
stars (Sect.~\ref{Decont_CMDs}). As a photometric quality constraint, only stars with 
\jj, \hh, and \ks\ errors lower than 0.1\,mag were used. Reddening corrections are 
based on the absorption relations $A_J/A_V=0.276$, $A_H/A_V=0.176$, $A_{K_S}/A_V=0.118$, 
and $A_J=2.76\times\ejh$ given by \citet{DSB2002}, with $R_V=3.1$, considering the 
extinction curve of \citet{Cardelli89}. The reddening values are derived from the
2MASS CMDs (Sect.~\ref{DFP}).

\subsection{Field decontamination}
\label{Decont_CMDs}

Field-star decontamination is usually required for the proper identification and 
characterisation of star clusters, especially those near the Galactic equator and/or 
with important fractions of faint stars. Cr\,197 and vdB\,92 are located in the $3^{rd}$ 
Galactic quadrant (Table~\ref{tab1}), which makes field-star contamination a minor issue 
(e.g. \citealt{ProbFSR}). However, their CMDs are dominated by PMS stars (Sect.~\ref{DFP}) 
and, thus, decontamination is important to avoid confusion with the red dwarfs of the 
Galactic field.

A summary of different decontamination approaches is provided in \citet{N2244}. In this 
paper we apply the decontamination algorithm developed in \citet{BB07}, together with an 
improvement that is described below. 

\begin{figure*}
\begin{minipage}[b]{0.50\linewidth}
\includegraphics[width=\textwidth]{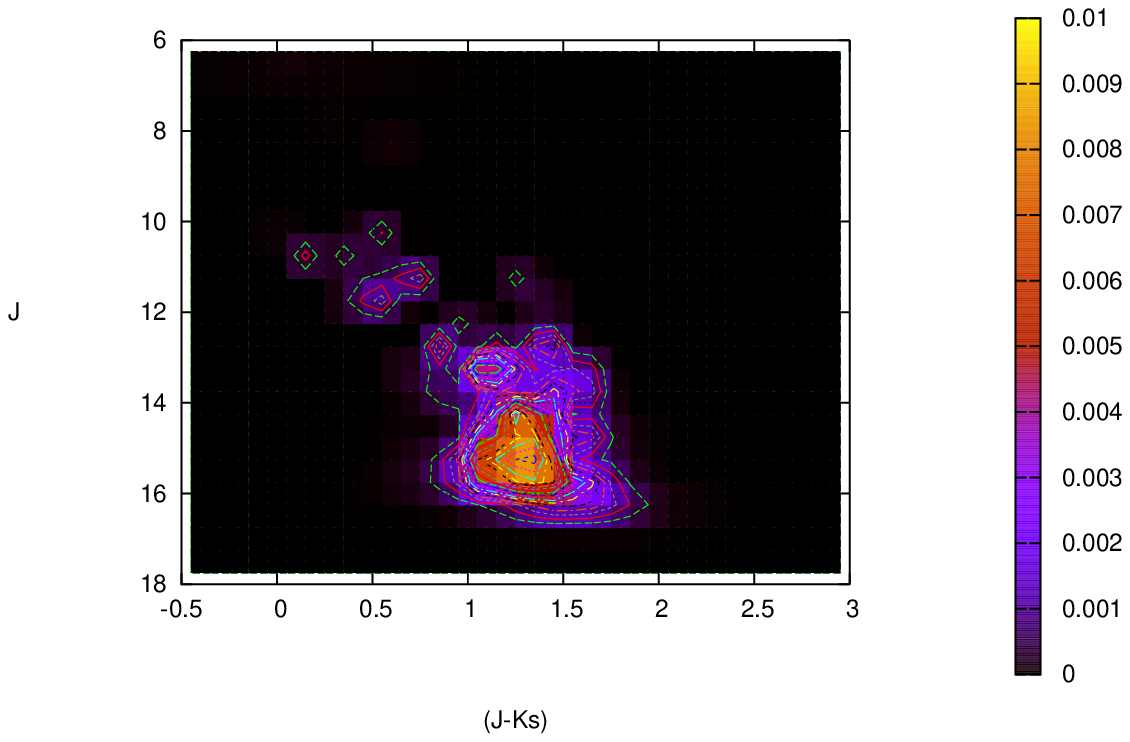}
\end{minipage}\hfill
\begin{minipage}[b]{0.50\linewidth}
\includegraphics[width=\textwidth]{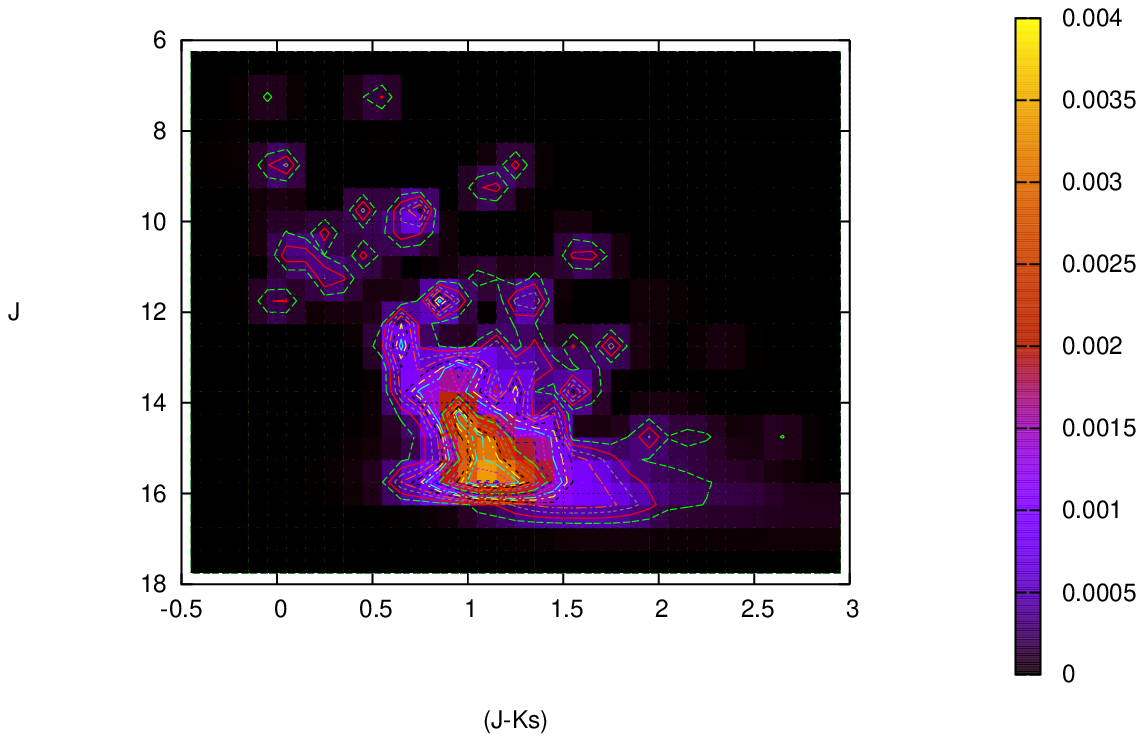}
\end{minipage}\hfill
\begin{minipage}[b]{0.50\linewidth}
\includegraphics[width=\textwidth]{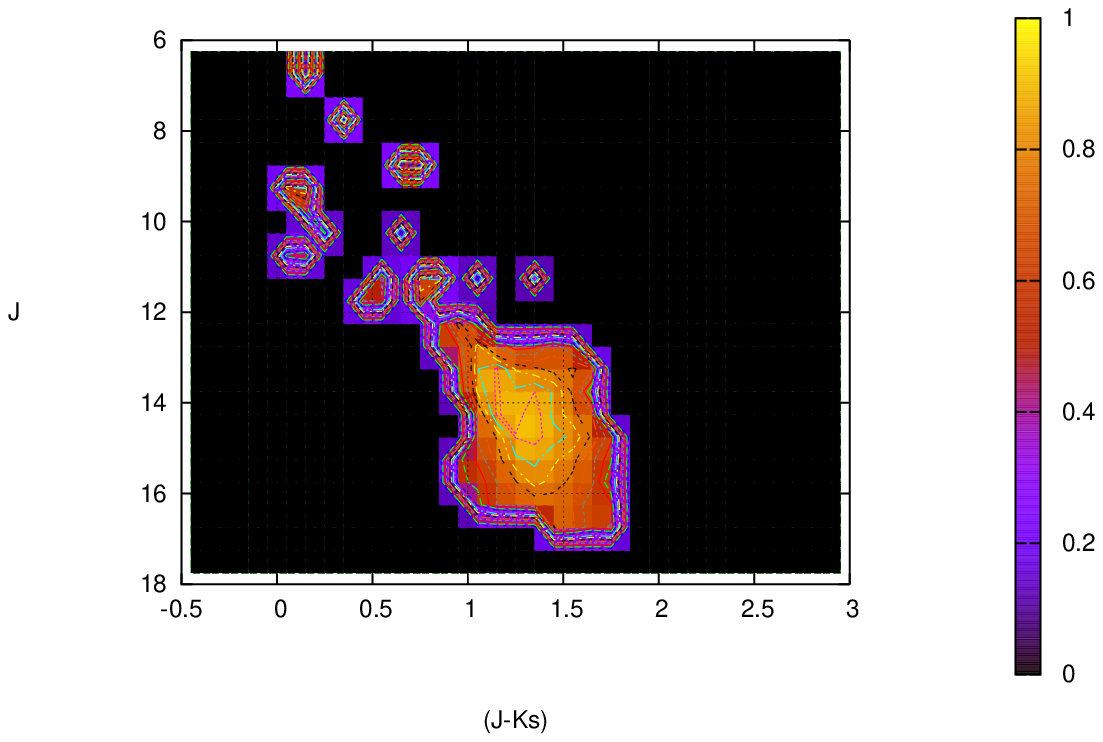}
\end{minipage}\hfill
\begin{minipage}[b]{0.50\linewidth}
\includegraphics[width=\textwidth]{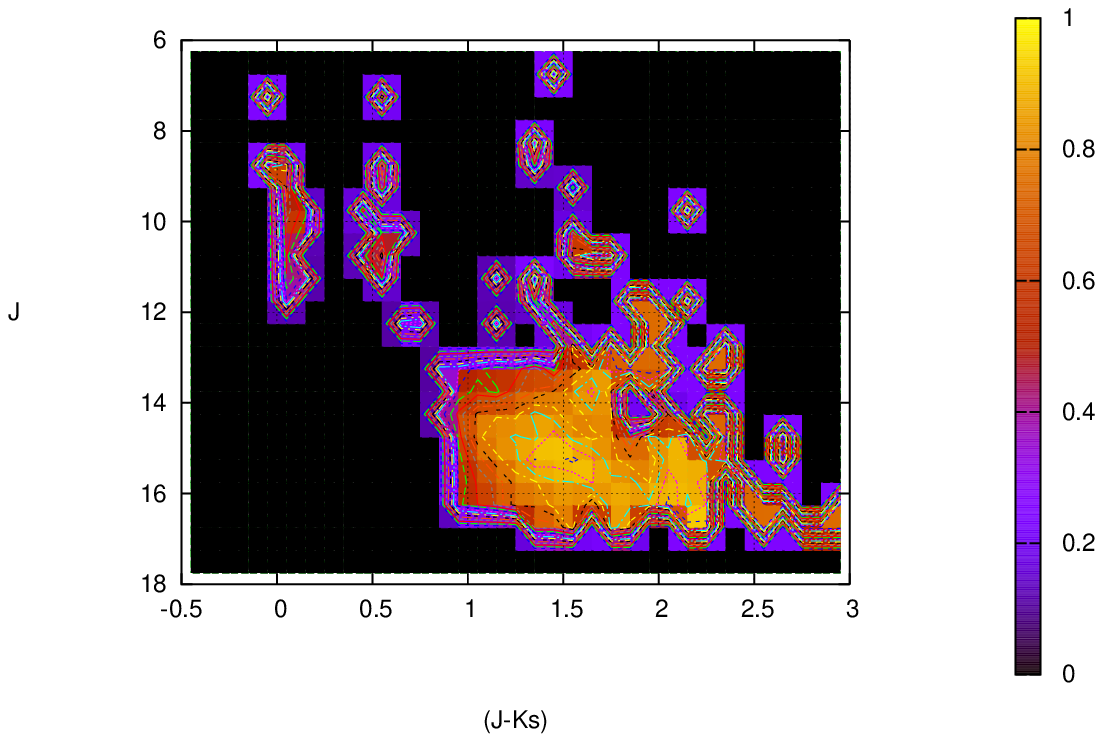}
\end{minipage}\hfill
\caption[]{Top panels: number-density ($\eta_{mem}$, in $\rm stars\,arcmin^{-2}$) of the 
probable member stars present in the decontaminated CMDs extracted within $R=10\arcmin$ of 
Cr\,197 (left) and $R=15\arcmin$ of vdB\,92 (right). Bottom: survival frequency of the 
decontaminated stars.}
\label{fig2}
\end{figure*}

Photometric uncertainties are explicitly taken into account, in the sense that stars
have a non-negligible probability (assumed to be Gaussian) of having a magnitude and
colour significantly different from the average values. The algorithm {\em (i)} divides 
the full range of magnitude and colours of a CMD into a 3D grid of cells with axes along 
the \jj\ magnitude and the \jh\ and \jk\ colours, {\em (ii)} computes the total (i.e. 
member$+$field) number-density of stars within a given cell ($\eta_{tot}$), 
{\em (iii)} estimates the number-density of field stars ($\eta_{fs}$) based on the number 
of comparison field stars with similar magnitude and colours as those in the cell, {\em (iv)}
computes the expected number-density of member stars, $\eta_{mem}=\eta_{tot}-\eta_{fs}$, 
{\em (v)} converts the number-density $\eta_{fs}$ into the estimated number of field 
stars and subtracts it from each cell, and {\em (vi)} after subtraction of the field stars, 
the remaining $N^{cell}_{clean}$ stars in each cell are identified for further use (see 
below). The wide annulus between $50\arcmin<R<80\arcmin$ (outside the cluster radius - 
Table~\ref{tab2}) is used as the comparison field for Cr\,197; for vdB\,92 it is located 
between $30\arcmin<R<80\arcmin$. The initial cell dimensions are $\Delta\jj=1.0$ and
$\Delta\jh=\Delta\jk=0.2$, but cell 
sizes half and twice the initial values are also used. As a new feature, we also consider 
shifts in the grid positioning by $\pm1/3$ of the respective cell size in the 2 colours and 
magnitude axes. When all the different grid/cell size setups are applied, the decontamination 
takes into account 729 independent combinations. 

Each setup produces a total number of member stars, $N_{mem}=\sum_{cell}N^{cell}_{clean}$,
from which we compute the expected total number of member stars $\left<N_{mem}\right>$ by 
averaging out $N_{mem}$ over all combinations. Stars (identified in step {\em (vi)} above) 
are ranked according to the number of times they survive all runs (survival frequency), and 
only the $\left<N_{mem}\right>$ highest ranked stars are considered cluster members and 
transposed to the respective decontaminated CMD (e.g. Figs.~\ref{fig5} and \ref{fig6}). 

The subtraction efficiency, i.e. the difference between the expected number of field stars 
(which may be fractional) and the number of stars effectively subtracted (which is integer) 
from each cell, summed over all cells, is 99.4\% for vdB\,92 and 96.7\% for Cr\,197. As a 
caveat we note that the present decontamination approach implicitly assumes that the
field colour-magnitude distribution somewhat matches that of the cluster. Figure~\ref{fig1}
shows some gas and dust around both clusters, which might lead to appreciable reddening and
differential reddening. While the effect in the foreground contamination may be small, it 
should be more important in the background. However, the $3^{rd}$ Galactic quadrant 
location of Cr\,197 and vdB\,92 is expected to minimise the background contamination.

The number-density of the probable member stars ($\eta_{mem}$) are shown in the $\jj\times\jk$ 
CMDs in Figure~\ref{fig2} (top panels). In both cases, most of the stars are relatively faint 
and red, covering a wide range in colour, which is expected of PMS stars somewhat affected by 
differential reddening. Also shown in Figure~\ref{fig2} (bottom panels) is the survival 
frequency of the decontaminated stars. Again, the highest membership probabilities occur 
among the PMS stars. 

\subsection{Decontaminated surface density maps}
\label{DecOut}

Our decontamination approach relies upon differences in the colour and magnitude 
distribution of stars located in separate spatial regions. For a star cluster, which can 
be characterised by a single-stellar population projected against a Galactic stellar field,
the decontaminated surface-density is expected to present a marked excess at the 
assumed cluster position.

\begin{figure}
\resizebox{\hsize}{!}{\includegraphics{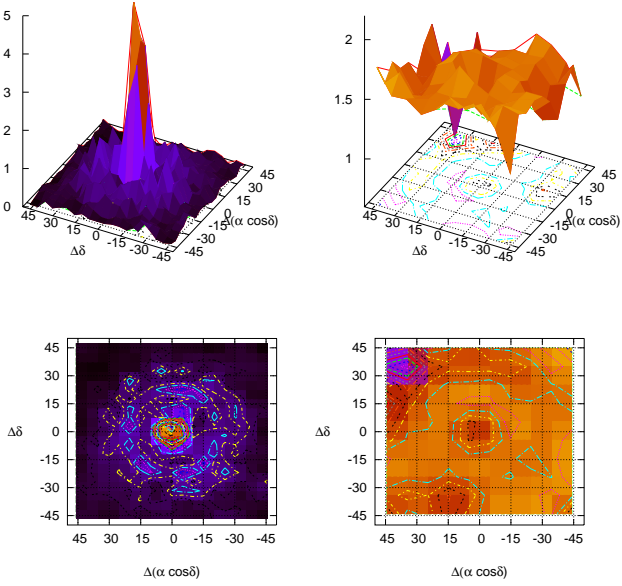}}
\caption[]{Left: Stellar surface-density $\sigma(\rm stars\ arcmin^{-2})$ computed with
field-decontaminated photometry to enhance the cluster/background contrast of Cr\,197. 
Right: Average \jk\ colour. $\Delta(\alpha~\cos(\delta))$ and $\Delta\delta$ in arcmin.}
\label{fig3}
\end{figure}

\begin{figure}
\resizebox{\hsize}{!}{\includegraphics{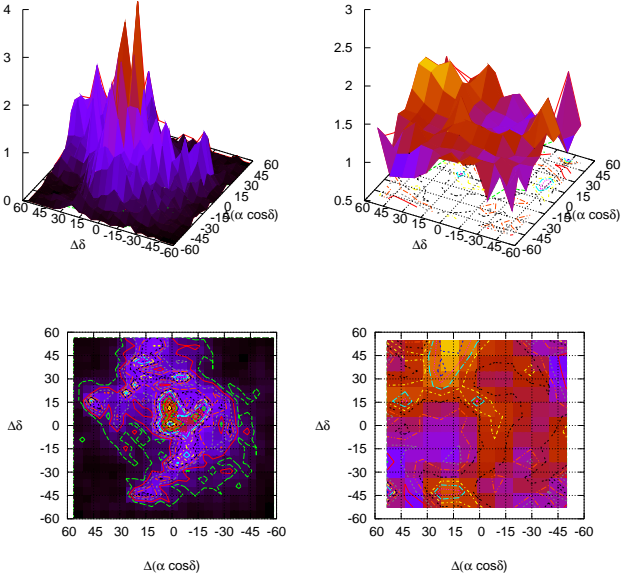}}
\caption[]{Same as Fig.~\ref{fig3} for vdB\,92.}
\label{fig4}
\end{figure}

Maps of the spatial distribution of the stellar surface-density ($\sigma$, in units of 
$\rm stars\,arcmin^{-2}$), built with field-star decontaminated photometry to maximise 
the cluster/background contrast, are shown in Figs.~\ref{fig3} (Cr\,197) and \ref{fig4} 
(vdB\,92). Also shown are the isopleths, in which cluster size and geometry can be 
observed. The surface density is computed in a rectangular mesh with cells 
$2.5\arcmin\times2.5\arcmin$ wide, reaching total offsets of 
$|\Delta\alpha~\cos(\delta)|=|\Delta\delta|\approx45\arcmin$ with respect to the cluster 
centre (Table~\ref{tab1}).

Despite the gas and dust associated with the present clusters (Fig.~\ref{fig1}), the central 
excesses show up markedly in the decontaminated surface-density distributions. Cr\,197 
has a well-defined, approximately round, and narrow stellar distribution (Fig.~\ref{fig3}, 
left panel). Its \jk\ colour distribution is rather uniform, within $1.0\la\jk\la1.8$, with 
a blueward dip, due to the few bright and rather blue stars, about the centre (right panel). 
The stellar distribution of vdB\,92, on the other hand, presents a broad and elongated 
external region (Fig.~\ref{fig4}, left panel), with a complex colour distribution (right 
panel), within $0.8\la\jk\la2.5$. If most of these colour variations (with respect 
to the average) are due to non-uniform reddening - and not to systematic differences in 
the stellar content - the upper limits to the differential reddening distribution would be 
$\Delta\aV\la2.5$\,mag for Cr\,197 and $\Delta\aV\la5$\,mag for vdB\,92. 

\section{Fundamental parameters}
\label{DFP}

\begin{figure}
\resizebox{\hsize}{!}{\includegraphics{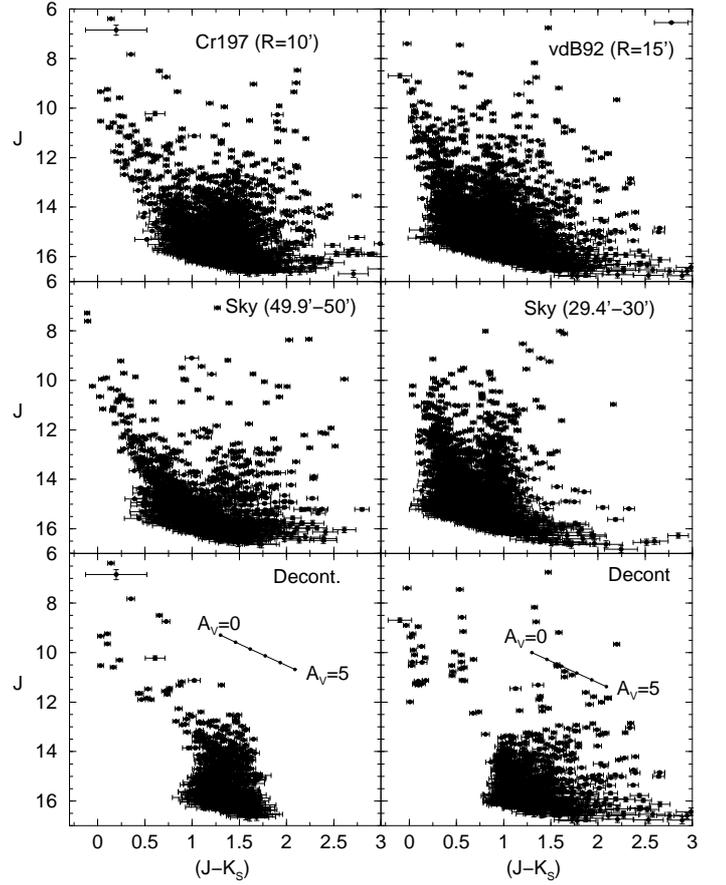}}
\caption[]{$\jj\times\jk$ CMDs of Cr\,197 (left) and vdB\,92 (right) showing the
observed photometry for representative regions (top) and the equal-area comparison 
fields (middle). The decontaminated CMDs are shown in the bottom panels. Reddening 
vectors for $\aV=0-5$ are shown in the bottom panels.}
\label{fig5}
\end{figure}

$\jj\times\jk$ CMDs built with the raw photometry of Cr\,197 and vdB\,92 are shown 
in the top panels of Fig.~\ref{fig5}. In both cases, the sampled region contains 
most of the cluster stars (Fig.~\ref{fig8}). Features typical of very young OCs,
such as a relatively vertical and poorly-populated MS, together with a large 
population of faint and red PMS stars, can be seen when these CMDs are compared
to those extracted from the equal-area offset fields\footnote{The equal-area field 
extractions are used only for qualitative comparisons, since the decontamination 
uses a wide surrounding ring area (Sect.~\ref{Decont_CMDs}).} (middle panels).

\begin{figure}
\resizebox{\hsize}{!}{\includegraphics{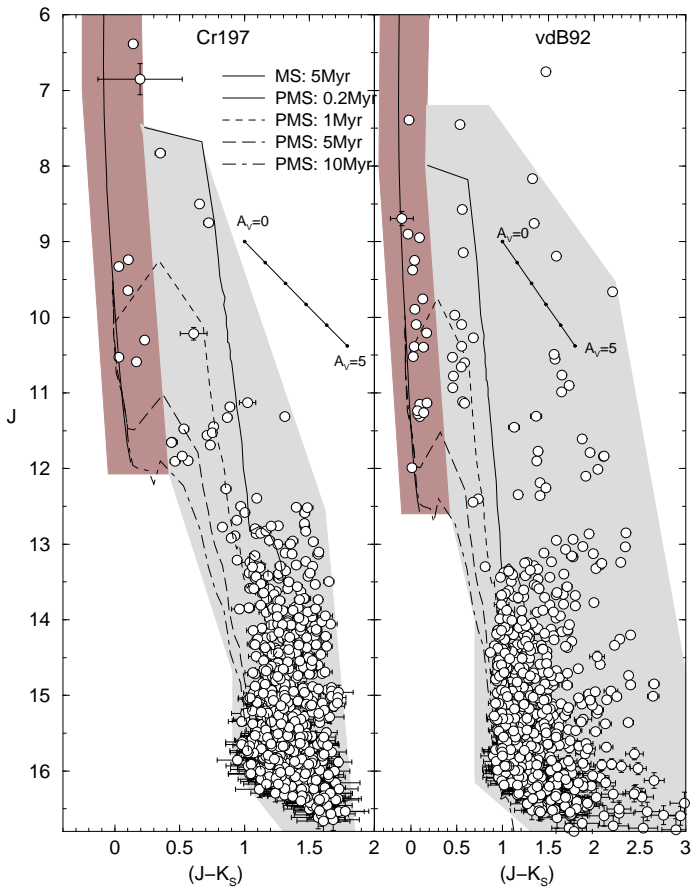}}
\caption[]{Adopted MS$+$PMS isochrone solutions to the decontaminated CMDs. 
Reddening vectors as in Fig.~\ref{fig5}. Shaded polygons show the MS (dark-gray) 
and PMS (light-gray) colour-magnitude filters (Sect.~\ref{struc}).}
\label{fig6}
\end{figure}

As expected, the decontaminated CMDs (bottom panels of Fig.~\ref{fig5}) contain 
essentially the typical stellar sequences of mildly reddened young clusters, with 
a developing MS and a significant population of PMS stars. Also, in both cases
the colour distribution at faint magnitudes ($\jj\ga13$) is wider than the spread 
predicted purely by PMS models (Fig.~\ref{fig6}), which implies internal differential 
reddening. To examine this issue we include in Figs.~\ref{fig5}-\ref{fig6} reddening 
vectors computed with the 2MASS ratios (Sect.~\ref{2mass}) for visual absorptions 
$\aV=0~{\rm to}~5$. Taking into account the PMS isochrone fit (Fig.~\ref{fig6}), the 
differential reddening appears to be lower than $\Delta\aV=5$, especially for Cr\,197,
which is consistent with the values estimated in Sect.~\ref{DecOut}.

The fundamental parameters are derived by means of the constraints provided by the 
field-decontaminated CMD morphologies, especially the combined MS and PMS star 
distribution (Fig.~\ref{fig6}). We adopt solar metallicity isochrones because both 
objects are young and located not far from the Solar circle (see below), a region 
essentially occupied by $[Fe/H]\approx0.0$ OCs (\citealt{Friel95}). Padova isochrones 
(\citealt{Girardi2002}) computed with the 2MASS \jj, \hh, and \ks\ 
filters\footnote{{\em http://stev.oapd.inaf.it/cgi-bin/cmd}. These isochrones are 
very similar to the Johnson-Kron-Cousins ones (e.g. \citealt{BesBret88}), with 
differences of at most 0.01\,mag in colour (\citealt{TheoretIsoc}). } are used to 
represent the MS. Isochrones of \citet{Siess2000} with ages in the range 0.2---10\,Myr 
are used to characterise the PMS sequences. Since the decontaminated CMD morphologies 
are very similar and typical of young ages (Fig.~\ref{fig6}), a similar isochrone solution 
is expected to apply to both objects.

As summarised in \citet{NJ06}, sophisticated approaches are available for analytical
CMD fitting, especially the MS. However, given the poorly-populated MSs, the 2MASS 
photometric uncertainties for the fainter stars, the important population of PMS stars, 
and the differential reddening, we applied the direct comparison of isochrones with 
the decontaminated CMD morphology. The fits are made {\em by eye}, taking the combined 
MS and PMS stellar distributions as constraint, allowing as well for variations due
to photometric uncertainties (which, given the restrictions imposed in Sect.~\ref{2mass}, 
are small in both cases), and differential reddening. Specifically, we start with the 
MS$+$PMS isochrones set for zero distance modulus and reddening, and apply shifts in 
magnitude and colour to them. We take as {\em best-fit} the apparent distance modulus 
and reddening that simultaneously account for the blue border of the MS and PMS stellar
distributions. Both clusters present a significant fraction of stars redder than the 
youngest PMS isochrone. Most of this \jk\ excess towards the red is probably due to 
differential reddening. Below we discuss the results, which are shown in Fig.~\ref{fig6}. 

{\tt Cr\,197:}
Given the poorly-populated and nearly vertical MS, acceptable fits to the 
decontaminated MS are obtained with any isochrone with age in the range 1---10\,Myr.
If we take into account the differential
reddening, this age spread is consistent with the PMS stars, which are basically contained 
within the 0.2\,Myr and 10\,Myr isochrones as well. Thus, we take the 5\,Myr isochrone as 
a representative solution, and allow for a $\pm4$\,Myr age-spread in the star formation. 

\begin{figure}
\resizebox{\hsize}{!}{\includegraphics{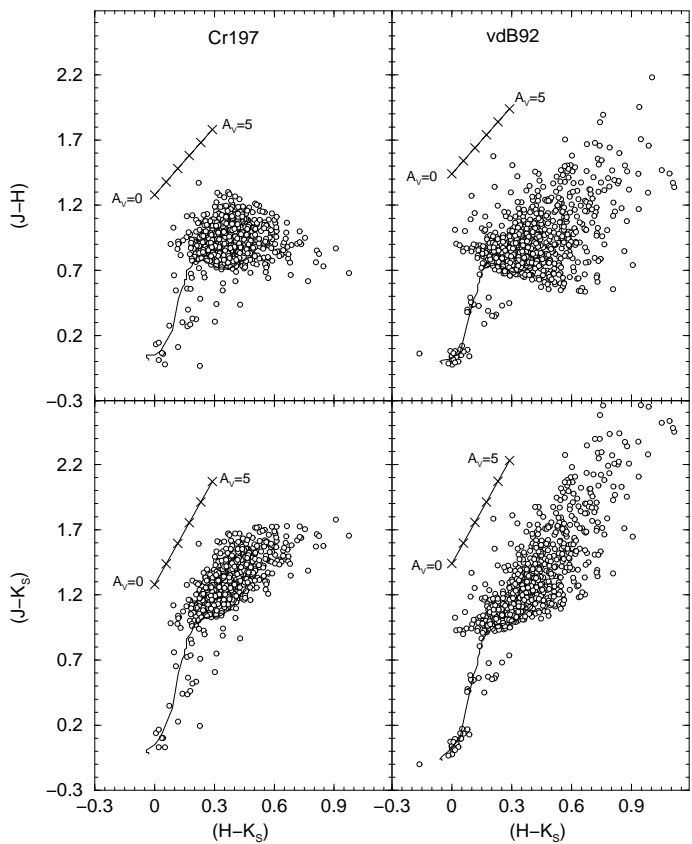}}
\caption[]{Colour-colour diagrams showing the decontaminated photometry and the 5\,Myr 
PMS track (\citealt{Siess2000}), set with the derived reddening values (Sect.~\ref{DFP}). 
MS stars lie to the blue side of the diagram. Error bars have been omitted for clarity. 
Reddening vectors as in Figs.~\ref{fig5} and \ref{fig6}.}
\label{fig7}
\end{figure}

With the adopted solution, the fundamental parameters of Cr\,197 are the near-IR reddening 
$\ejh=0.10\pm0.05$ ($\ebv=0.34\pm0.16$ or $A_V=1.0\pm0.5$), the observed and absolute distance 
moduli $\mMJ=10.4\pm0.4$ and $\mMo=10.11\pm0.42$, respectively, and the distance from the Sun 
$\ds=1.05\pm0.20$\,kpc. This distance is compatible with that obtained for the B4\,II star
HIP\,42908 ($B=7.65$ and $V=7.32$), $\ds=1.1$\,kpc. With $\rs=7.2\pm0.3$\,kpc (\citealt{GCProp}) 
as the Sun's distance to the Galactic centre\footnote{Derived by means of the Globular Cluster 
spatial distribution. Recently, \citet{Trippe08} found $\dgc=8.07\pm0.32$\,kpc while 
\citet{Ghez08} found $\dgc=8.0\pm0.6$\,kpc or $\dgc=8.4\pm0.4$\,kpc, under different 
assumptions.}, the Galactocentric distance of Cr\,197 is $\dgc=7.5\pm0.1$\,kpc, which puts 
it $\approx0.3$\,kpc outside the Solar circle. This solution is shown in Fig.~\ref{fig6}. 
The projected coordinate components are $(x_{GC},y_{GC},z_{GC})\approx(-7.4,-1.4,0.0)$ 
in kpc, which put Cr\,197 at the Orion-Cygnus and Sagittarius-Carina interarm region (e.g.
\citealt{Momany06}).
We also compute the integrated apparent and absolute magnitudes in the 2MASS bands for the 
member stars within $R\leq\rl$, together with the reddening corrected colours. The results
are $m_J=5.64\pm0.07$, $M_J=-4.76\pm0.12$, $\jh=0.19\pm0.09$, $\jk=0.32\pm0.09$.

{\tt vdB\,92:} 
Similarly to Cr\,197, any isochrone with age within 1---10\,Myr provides an
acceptable fit to the nearly vertical MS, while the PMS distribution is essentially 
contained within the
0.2---10\,Myr isochrones. With this solution, the fundamental parameters of vdB\,92 
are $\ejh=0.07\pm0.04$ ($\ebv=0.22\pm0.13$ or $A_V=0.7\pm0.4$), $\mMJ=10.9\pm0.4$, 
$\mMo=10.71\pm0.41$, $\ds=1.38\pm0.26$\,kpc, and $\dgc=8.3\pm0.2$\,kpc, thus 
$\approx1.0$\,kpc outside the Solar circle. Thus, with the coordinate components
$(x_{GC},y_{GC},z_{GC})\approx(-8.2,-1.0,-0.1)$, vdB\,92 approximately coincides
with the Orion-Cygnus arm  (e.g. \citealt{Momany06}). SIMBAD provides 5 stars with spectral type
and optical photometry to within $\approx10\arcmin$ of the central coordinates of vdB\,92
(Table~\ref{tab1}). These are HRW\,14 (B7\,III, $B=12.67$ and $V=12.09$), BD$-$11\,1763 
(B1.5\,V, $B=8.90$ and $V=8.91$), BD$-$11\,1761 (B2\,V, $B=9.32$ and $V=9.25$), HIP\,34133 
(B0\,V, $B=7.31$ and $V=7.34$), and NSV\,3364 (B3\,V, $B=9.55$ and $V=9.41$). The average
spectroscopic distance of these stars is $\ds=1.4\pm0.2$\,kpc, in excellent agreement with 
the photometric value for vdB\,92. For vdB\,92 we derive $m_J=6.35\pm0.03$, 
$M_J=-4.53\pm0.41$, $\jh=0.54\pm0.05$, $\jk=1.01\pm0.04$. Thus, Cr\,197 is intrinsically
brighter and somewhat bluer than vdB\,92.

Besides the similar MS age, Cr\,197 and vdB\,92 have in common a significant age 
spread ($\sim10$\,Myr) implied by the PMS stars, which indicates a non-instantaneous 
star formation. We found a similar scenario in our previous studies of, e.g. NGC\,4755, 
NGC\,6611 and NGC\,2244.

Although with a relatively low foreground absorption ($A_V\la1$), both cases present 
a significant fraction of PMS stars redder (and fainter) than predicted by 
the PMS tracks (Fig.~\ref{fig6}). Indeed, when transposed to near-IR colour-colour 
diagrams (Fig.~\ref{fig8}), the age and reddening solutions derived for Cr\,197 and 
vdB\,92 consistently match most of their field-star decontaminated photometry, but a 
significant fraction of the PMS stars appears to be very reddened. Most of the very 
red PMS stars occur along the respective reddening vectors. However, a small fraction 
appears to present an abnormal excess in \hk, especially vdB\,92, in the $\jh\times\hk$ 
diagram, which may come from PMS stars still bearing circumstellar discs and, thus, 
with excess in \hh\ (e.g. \citealt{Furlan09}; \citealt{N4755}). 

\section{Structural parameters}
\label{struc}

We derive structural parameters by means of the RDPs, which are the projected stellar 
number density around the cluster centre. Noise in the RDPs is minimised with 
colour-magnitude filters (Fig.~\ref{fig6}), which exclude stars with colours not 
compatible with those of the cluster\footnote{They are wide enough to include cluster 
MS and PMS stars, together with the photometric uncertainties and binaries (and other 
multiple systems).}. In previous works we have shown that this filtering procedure 
enhances the RDP contrast relative to the background, especially in crowded fields 
(e.g. \citealt{BB07}). 

\begin{figure}
\resizebox{\hsize}{!}{\includegraphics{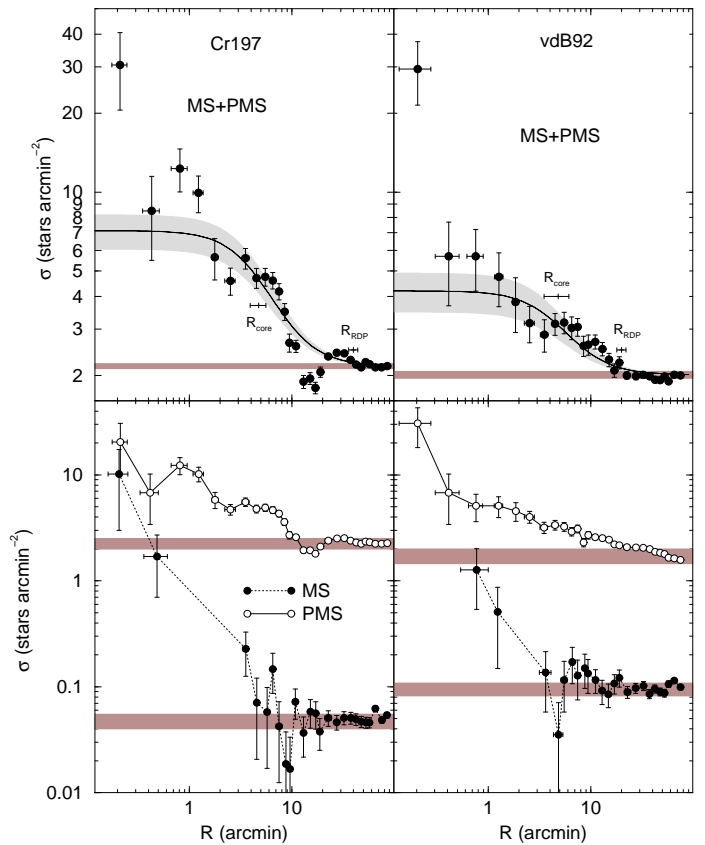}}
\caption[]{Top: Stellar RDPs built for the MS and PMS stars combined, together with
the best-fit King-like profile (solid line), the $1\sigma$ uncertainty (light-shaded
region) and the background level (shaded polygon). Note the pronounced central cusps.
Bottom: RDPs built isolately for the MS and PMS stars.}
\label{fig8}
\end{figure}

Rings of increasing width with distance from the cluster centre are used to preserve
spatial resolution along the full radial range. The set of ring widths used is 
$\Delta\,R=0.25,\ 0.5,\ 1.0,\ 2.5,\ {\rm and}\ 5\arcmin$, respectively for 
$0\arcmin\le R<0.5\arcmin$, $0.5\arcmin\le R<2\arcmin$, $2\arcmin\le R<5\arcmin$, 
$5\arcmin\le R<20\arcmin$, and $R\ge20\arcmin$. The residual background level of each 
RDP is the average number-density of field stars. The $R$ coordinate (and 
uncertainty) of each ring corresponds to the average position and standard deviation 
of the stars inside the ring. The resulting RDPs are shown in Fig.~\ref{fig8}. As 
a caveat we note that the projected stellar distribution of vdB\,92 is not smooth and
radially symmetric, especially at the outskirts (Fig.~\ref{fig4}). Thus, the use of 
circular rings to build the RDP might introduce some noise at large radii, but with 
little effect in the central parts. 

\begin{table*}
\caption[]{Derived structural parameters}
\label{tab2}
\renewcommand{\tabcolsep}{2.0mm}
\renewcommand{\arraystretch}{1.25}
\begin{tabular}{cccccccccccc}
\hline\hline
Cluster&$\sigma_{bg}$&$\sigma_0$&\rc&\rl&$\delta_c$&$1\arcmin$&$\sigma_{bg}$&$\sigma_0$&\rc&\rl\\
       &$\rm(*\,\arcmin^{-2})$&$\rm(*\,\arcmin^{-2})$&(\arcmin)&(\arcmin)& &(pc)&
$\rm(*\,pc^{-2})$&$\rm(*\,pc^{-2})$&(pc)&(pc)\\
(1)&(2)&(3)&(4)&(5)&(6)&(7)&(8)&(9)&(10)&(11)\\
\hline
Cr\,197$^\dagger$&$2.17\pm0.02$&$5.0\pm1.0$&$4.8\pm0.8$&$40.0\pm4.0$&$3.3\pm0.5$&0.305&$23.3\pm0.1$&$97.9\pm20.9$&
   $1.5\pm0.3$&$12.2\pm1.2$\\
   
Cr\,197$^\ddagger$&$2.15\pm0.04$&$4.7\pm1.0$&$5.2\pm0.9$&$40.0\pm4.0$&$3.2\pm0.5$&0.305&$23.1\pm0.2$&$92.8\pm17.9$&
   $1.6\pm0.3$&$12.2\pm1.2$\\
   
vdB\,92$^\dagger$&$2.01\pm0.06$&$2.2\pm0.7$&$4.8\pm1.3$&$20.0\pm2.0$&$2.1\pm0.3$&0.402&$12.5\pm0.4$&$13.6\pm4.4$&
   $1.9\pm0.5$&$8.0\pm0.8$\\
   
vdB\,92$^\ddagger$&$1.97\pm0.06$&$2.7\pm0.6$&$5.0\pm0.9$&$20.0\pm2.0$&$2.4\pm0.3$&0.402&$12.3\pm0.4$&$17.0\pm3.6$&
   $2.0\pm0.4$&$8.0\pm0.8$\\
   
\hline
\end{tabular}
\begin{list}{Table Notes.}
\item Col.~6: cluster/background density contrast parameter ($\delta_c=1+\sigma_0/\sigma_{bg}$)
computed with the King-like fit parameters. Col.~7: arcmin to parsec scale. ($\dagger$): background
level kept fixed; ($\ddagger$): background level allowed to vary.
\end{list}
\end{table*} 

We fit the RDPs with the function $\sigma(R)=\sigma_{bg}+\sigma_0/(1+(R/R_c)^2)$, 
where $\sigma_0$ and $\sigma_{bg}$ are the central and residual background stellar 
densities, and \rc\ is the core radius. Applied to star counts, this function is 
similar to that used by \cite{King1962} to the surface-brightness profiles in the 
central parts of globular clusters\footnote{Because of the relatively low number of 
stars in the present OCs, fluctuations in surface-brightness profiles are expected 
to be higher than those in RDPs. Alternative RDP fit functions are discussed in 
\citet{StrucPar}.}. To minimise degrees of freedom, only $\sigma_0$ and \rc\ are
derived from the fit, while $\sigma_{bg}$ is previously measured in the surrounding 
field and kept fixed.

The best-fit solutions (and uncertainties) are shown in Fig.~\ref{fig8}, and the 
corresponding structural parameters are given in Table~\ref{tab2}. In addition, we 
estimate the cluster radius (\rl), i.e. the distance from the centre where the cluster 
RDP and field fluctuations are statistically indistinguishable (e.g. \citealt{DetAnalOCs}), 
and the density contrast parameter $\delta_c=1+\sigma_0/\sigma_{bg}$ (Table~\ref{tab2}).
\rl\ may be taken as an observational truncation radius, whose value depends both on the 
radial distribution of member stars and the field density. 

Within uncertainties, the adopted King-like function provides a reasonable description
of the RDPs for $R\ga1\arcmin$ (Fig.~\ref{fig8}). The drop in the RDP of CR\,197 for
$13\la R(\arcmin)\la18$ is probably related to enhanced dust absorption. However, the RDP 
measured in the innermost region ($R\la0.3\arcmin$) in both cases presents a significant 
($\ga3\sigma$) excess, or cusp, over the fit. In old star clusters, this feature has been 
attributed to a post-core collapse structure, as detected in some globular clusters (e.g. 
\citealt{TKD95}). Some Gyr-old OCs, e.g. NGC\,3960 (\citealt{N3960}) and LK\,10 
(\citealt{LKstuff}), also display such a dynamical evolution-related feature. Interestingly,
this cusp also occurs in the RDP of some very young ($\la10$\,Myr) clusters, e.g. NGC\,2244 
(\citealt{N2244}), NGC\,6823 (\citealt{Bochum1}), Pismis\,5 and NGC\,1931 (\citealt{Pi5}).
Such time-scales are too short for clusters to evolve into a post-core collapse. Thus, the 
cusp in young clusters is probably related to molecular cloud fragmentation and/or star
formation, and may suggest important early deviation from dynamical equilibrium 
(Sect.~\ref{Discus}). As a dynamically-related alternative explanation for the 
RDP cusps, we note that numerical simulations of clusters that form dynamically cool 
(subvirial) and with a fractal structure (\citealt{Allison09}), suggest that mass 
segregation may occur on timescales comparable to their crossing times, typically a 
few Myr. Also, clusters that form with significant substructure will probably develop 
an irregular central region, unless such a region collapses and smooths-out the initial substructure.

Finally, we also show in Fig.~\ref{fig8} (bottom panels) the RDPs built after
isolating the MS and PMS stars by means of the respective colour-magnitude filters 
(Fig.~\ref{fig6}). In both clusters, MS stars are more concentrated than the
PMS ones and, as expected, the fraction of PMS stars (with respect to the total
number) increases with cluster radius, and largely dominates over the MS stars 
especially for $R\ga5\arcmin$. The innermost region of vdB\,92 is occupied essentially 
by PMS stars. The different residual background levels between the PMS and MS RDPs
reflect the dominant presence of faint and red stars in the field, with respect to 
the bright and blue ones. As a consequence, the density-contrast parameter of the
MS stars in Cr\,197 is $\approx20$ higher than that of the PMS stars. The absence of
MS stars in the innermost RDP bin precludes this comparison for vdB\,92.

Considering the fit parameters (Table~\ref{tab2}), the density-contrast parameter is 
relatively low in both cases, $2.1\la\delta_c\la3.3$. However, when the measured 
innermost RDP value is considered (Fig.~\ref{fig8}), it increases to $\delta_c=15.1\pm4.6$ 
and $\delta_c=15.7\pm4.0$, respectively for Cr\,197 and vdB\,92. The latter values are 
more representative of the visual contrast produced by the few relatively bright stars 
in the central regions of both objects.

Alternatively, we fitted the RDPs allowing for variations of the background level.
Within the uncertainties, the corresponding parameters (Table~\ref{tab2}) are compatible
with the previous (fixed background level) ones. Taken at face values, the core radii - 
derived from the King-like fits - of Cr\,197 ($\rc\approx1.5$\,pc) and vdB\,92 
($\rc\approx2.0$\,pc) are somewhat larger than the median value of the \rc\ distribution 
derived for a sample of relatively nearby OCs by \citet{Piskunov07}. 

\section{Stellar mass estimate}
\label{MF}

By far, the CMDs of Cr\,197 and vdB\,92 are dominated in number by PMS stars, followed 
by the developing, poorly-populated MS (Fig.~\ref{fig6}), which implies that most of
the cluster mass is stored in the PMS stars. To compute the cluster masses we work 
with the field-decontaminated photometry within $R\le\rl$ and count the number of 
stars that belong either to the MS or PMS. 

For the MS, the corresponding mass of each star is taken from the mass-luminosity 
relation derived from the isochrone solution (Sect.~\ref{DFP}). The MS mass range, 
in both cases, lies within $\approx2\,\ms - 30\,\ms$. Summing up these
values for all stars produces the total number ($n_{MS}$) and mass ($m_{MS}$) of 
MS stars. Because of the differential reddening, individual masses cannot be
assigned to the PMS stars. Thus, we simply count the number of PMS stars and adopt 
an average mass value to estimate $n_{PMS}$ and $m_{PMS}$. To compute the average
PMS mass value we use the \citet{Kroupa2001} initial mass function\footnote{Defined
as $dN/dM\propto m^{-(1+\chi)}$, it assumes the slopes $\chi=0.3\pm0.5$ for 
$0.08\leq m(\ms)\leq0.5$, $\chi=1.3\pm0.3$ for $0.5\leq m(\ms)\leq1.0$, and 
$\chi=1.35$ for $m(\ms)>1.0$.} for PMS masses between $0.08\,\ms - 7\,\ms$. The
result is $\bar{m}_{PMS}=0.6\,\ms$. Taking into account the uncertainty in 
\rl\ (Table~\ref{tab2}), we obtain for Cr\,197 $n_{MS}=26^{+10}_{-7}$,
$m_{MS}=172^{+66}_{-46}\,\ms$, $n_{PMS}=809^{+170}_{-62}$, and 
$m_{PMS}=485^{+78}_{-37}\,\ms$. Thus, Cr\,197 contains $\approx835$ stars and
$\mcl\approx660^{+102}_{-59}\,\ms$ of stellar mass. The values for vdB\,92 are
$n_{MS}=13^{+8}_{-5}$, $m_{MS}=106^{+65}_{-41}\,\ms$, $n_{PMS}=1067^{+130}_{-50}$, 
and $m_{PMS}=640^{+78}_{-30}\,\ms$. The cluster vdB\,92 thus contains $\approx1080$ 
stars and $\mcl\approx750^{+101}_{-51}\,\ms$ of stellar mass. 

Both objects present similar values of member MS and PMS stars, and total mass. Besides, 
as anticipated by the CMDs (Fig.~\ref{fig6}), the MS is poorly-populated in each case, to
the point that the mass stored in the PMS stars is the dominant ($74\%$ in Cr\,197 and 
$85\%$ in vdB\,92) component of the cluster mass. However, these masses may still be
somewhat higher, since because of the differential reddening and the 2MASS photometric 
limit, we may not detect the very-low mass PMS stars. Also, given the presence of some 
dust and gas in Cr\,197 and vdB\,92 (Fig.~\ref{fig1}), their total cluster masses may 
be a bit higher than the present estimates. Finally, the fact that most of the mass 
is stored in the faint PMS stars may explain the significant discrepancies with respect 
to previous mass determinations made in the optical (Sect.~\ref{RecAdd}).

\section{Proper motions}
\label{propMot}

Given the relative proximity of both clusters (Table~\ref{tab1}), we can use
proper motion data to probe kinematical properties of the member stars. For this
purpose we use the Third U.S. Naval Observatory CCD Astrograph Catalog 
(UCAC3, \citealt{Zach09}), which also provides the 2MASS photometry for each
star. 

Proper motions were obtained for the same central coordinates and extraction
radius as those used to extract the 2MASS photometry (Table~\ref{tab1}). The
region analysed is contained within $R=10\arcmin$, both for Cr\,197 and
vdB\,92, with the comparison field located within $60\arcmin - 80\arcmin$. 
Also, we applied the respective colour-magnitude filters (Sect.~\ref{struc}) 
before computing the proper motion distributions. Conversion from $\rm mas\,yr^{-1}$ 
to \kms\ was based on the respective cluster distances (Table~\ref{tab1}).

\begin{figure}
\resizebox{\hsize}{!}{\includegraphics{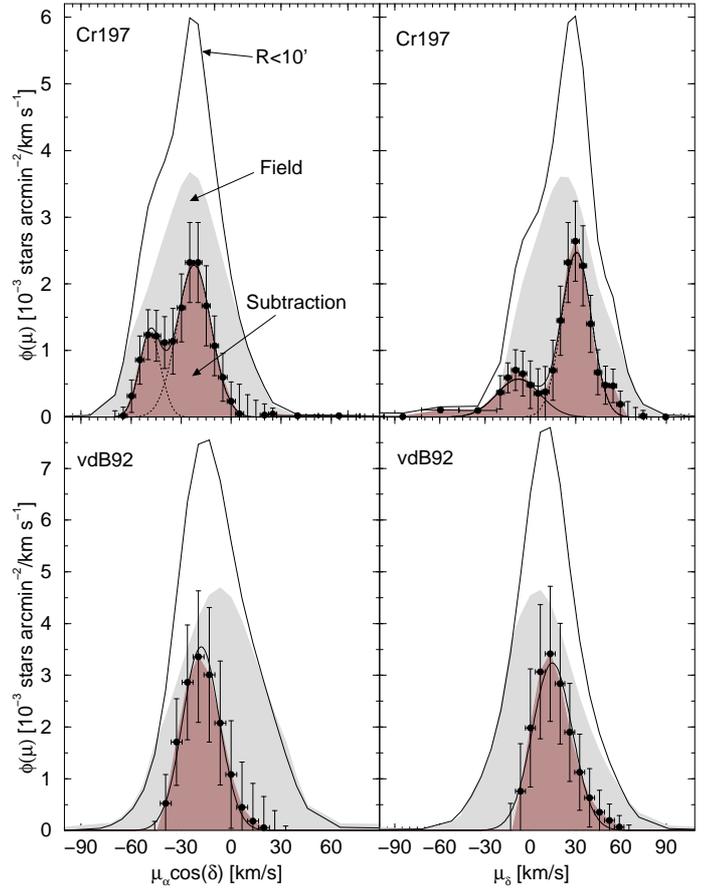}}
\caption[]{Proper motion distribution of Cr\,197 (top) and vdB\,92 (bottom). In both 
cases, the field contribution (light-gray) is subtracted from that of the $R<10\arcmin$
region (heavy-solid line) to produce the intrinsic distribution (heavy-gray). Gaussians
are fitted to the resulting distributions (see text for details). For clarity, error 
bars are only shown in the subtracted profiles.}
\label{fig9}
\end{figure}

The proper motion distributions ($\phi(\mu)=dN/d\mu$, in units of number of stars
per arcmin$^2$ per \kms), for both the $R=10\arcmin$ and field regions, are shown 
in Fig.~\ref{fig9}. The intrinsic distribution corresponds to the subtraction
of the field contribution to the $R=10\arcmin$ region. 

Cr\,197 appears to require two components of similar velocity dispersion to describe 
its intrinsic proper motion distribution, while a single one applies to vdB\,92. The 
best-fits, obtained with gaussians, are shown in Fig.~\ref{fig9}, and the corresponding
parameters are given in Table~\ref{tab_PM}. In velocity space, the 2 proper motion 
distributions of Cr\,197 have peaks separated by $\approx26\,\kms$ and $\approx39\,\kms$,
respectively in right ascension and declination. 

\begin{table}
\caption[]{Gaussian fit parameters}
\label{tab_PM}
\renewcommand{\tabcolsep}{0.65mm}
\renewcommand{\arraystretch}{1.25}
\begin{tabular}{lccccc}
\hline\hline
Cluster&$\overline{\mu_\alpha\,\cos(\delta)}$&$\sigma_\alpha$&$\overline{\mu_\delta}$&
$\sigma_\delta$&$\sigma_v$\\
       &(\kms)&(\kms)&(\kms)&(\kms)&(\kms)\\
\hline
Cr\,197&$-22.2\pm4.1$&$10.2\pm3.7$&$-8.1\pm4.0$&$14.4\pm4.3$&$22\pm5$\\
   
Cr\,197&$-48.5\pm5.0$&$6.6\pm3.6$&$+30.6\pm2.1$&$9.9\pm1.8$&$15\pm3$\\
   
vdB\,92&$-18.0\pm1.6$&$11.1\pm1.3$&$+14.5\pm2.1$&$13.1\pm1.7$&$21\pm2$\\
   
\hline
\end{tabular}
\begin{list}{Table Notes.}
\item $\sigma_\alpha$ and $\sigma_\delta$ are the velocity dispersions in
right ascension and declination, respectively. Assuming spherical symmetry, 
we define $\sigma_v^2=\frac{3}{2}(\sigma_\alpha^2+\sigma_\delta^2)$.
\end{list}
\end{table}

The spatial velocity dispersion of a cluster that is in virial equilibrium can be 
expressed as (\citealt{Spitzer87}) $\sigma^2 = \frac{G\,M_D}{\eta\,R_{eff}} \approx 
0.5\left(\frac{M_D}{10^3\ms}\right)\left(\frac{R_{eff}}{1\,pc}\right)^{-1}\,(\kms)^2 $, 
where $G$ is the gravitational constant, $M_D$ is the dynamical mass (assumed to be 
stored only in stars), $\eta\approx9.75$ is a constant (that roughly depends on the 
density profile), and $R_{eff}$ is the effective, or half-light, radius. Thus, 
clusters with $M_D\sim10^3\ms$ and $R_{eff}\sim1$\,pc are expected to present 
$\sigma\sim1\,\kms$. 

In the last column of Table~\ref{tab_PM} we provide an estimate of the spatial velocity 
dispersion ($\sigma_v$), computed assuming spherical symmetry for Cr\,197 and vdB\,92,
$\sigma_v^2=\frac{3}{2}(\sigma_\alpha^2+\sigma_\delta^2)$. Clearly, both clusters
present $\sigma_v$ values much higher than the expected dynamical ones\footnote{Because 
the velocity dispersions considered here are derived from proper motions, the binaries
contribution to $\sigma_v$ (\citealt{GSPZ10}) is expected to be minimal.}. Such a
super-virial state has been linked to the impulsive expulsion of the parental molecular 
cloud residual gas, driven essentially by the strong winds of massive stars and 
supernovae (\citealt{GoBa06}), which leads to the high rate of dissolution of the
young clusters (e.g. \citealt{LL2003}). A similar scenario has been observed to occur 
in the dissolving OC NGC\,2244, which has a spatial velocity dispersion of 
$\sigma_v\approx35\,\kms$ (\citealt{CGZ07}), which by far exceeds the expected 
dynamical value (\citealt{GKG08}).

\section{Discussion} 
\label{Discus}

In previous sections we derived a set of astrophysical parameters for Cr\,197 and vdB\,92.
We now compare them with the same parameters derived (using the same methods and
photometry) for a sample of {\em classical} OCs, basically gravitationally bound
clusters, characterised by a range of properties (e.g. age, mass, size, etc), and 
located in different environments. The template sample contains some relatively 
nearby and bright OCs (\citealt{DetAnalOCs}; \citealt{N4755}) together with a few
projected towards the central Galaxy (\citealt{BB07}). The young OCs NGC\,6611
(\citealt{N6611}), NGC\,6823 (\citealt{Bochum1}) and NGC\,2239 (\citealt{N2244}) 
are included for comparison with gravitationally bound objects of similar ages, 
while the young NGC\,2244 (\citealt{N2244}) might be dynamically evolving into 
an OB association. 

\begin{figure}
\resizebox{\hsize}{!}{\includegraphics{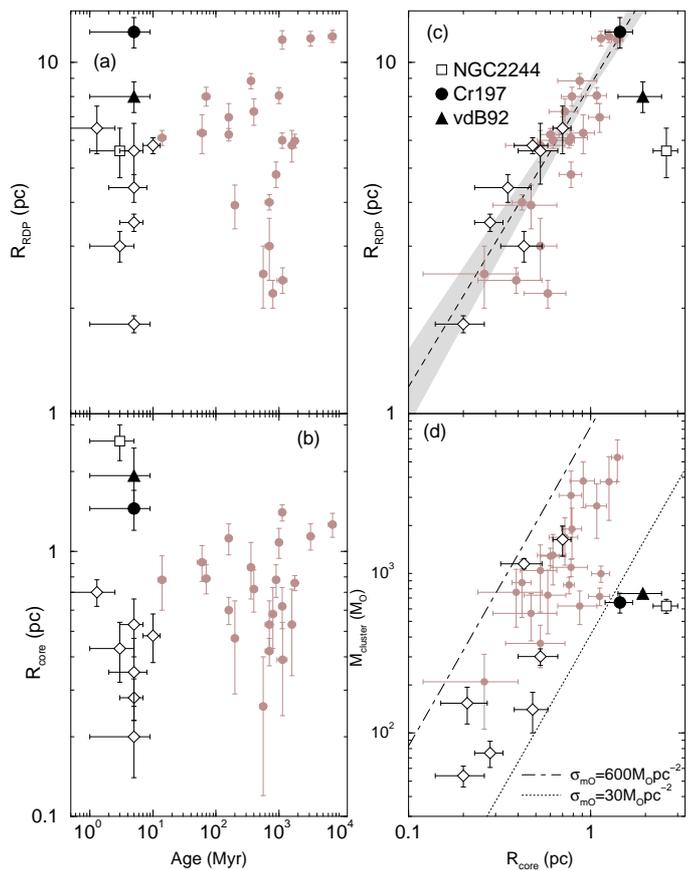}}
\caption{Diagrams dealing with astrophysical parameters of OCs. Gray-shaded circles:
template OCs. The young OC NGC\,2244 is indicated for comparison purposes. The 
analytical relations in panels (c) and (d) are discussed in the text.}
\label{fig10}
\end{figure}

The comparisons are based on diagnostic-diagrams built with template clusters 
(Fig.~\ref{fig10}). Panels (a) and (b) deal with the dependence of cluster (\rl) and 
core (\rc) radii on age, 
respectively. In terms of \rl, vdB\,92, and especially Cr\,197, present values higher 
than the other young OCs in the sample. Cr\,197 is as big as some Gyr-old OCs. With 
respect to \rc, both objects have a core significantly larger than that of the
{\em classical} (young and old) OCs, comparable to the core of the disrupting 
NGC\,2244. Within uncertainties, Cr\,197 follows the relation 
$\rl=(8.9\pm0.3)\times R_{\rm C}^{(1.0\pm0.1)}$ (panel c), derived with the 
template sample. Similarly to NGC\,2244, vdB\,92, on the other hand, is an outlier.

For gravitationally bound star clusters in which the projected mass density follows 
a King-like profile (e.g. \citealt{StrucPar}) characterised by a central surface 
mass-density $\sigma_{M0}$, the mass inside \rl\ can be expressed as a function of 
\rc, $\sigma_{M0}$, and the ratio \rl/\rc, as $\rm M_{clus}=\pi\,R^2_{C}\sigma_{M0}\ln\left
[1+\left(\rl/\rc\right)^2\right]$. With the relation between \rc\ and \rl\ implied 
by panel (c), this equation becomes $\rm M_{clus}\approx13.8\sigma_{M0}\,R^2_{C}$. 
Thus, the template OCs are constrained within King-like distributions with 
$\rm30\la\sigma_{M0}\,(\ms\,pc^{-2})\la600$ (panel d). Cr\,197 and vdB\,92 
(together with NGC\,2244) are outliers with respect to the distribution of bound 
OCs in the plane $R_{\rm C}\times M_{clus}$, in the sense that their locii imply 
exceedingly low central mass densities ($\rm4\la\sigma_{M0}\,(\ms\,pc^{-2})\la30$). 
Significantly smaller core radii and/or higher cluster mass would be required for 
them to be contained within the {\em classical} OC boundaries. 

Together with the kinematical analysis (Sect.~\ref{propMot}), a possible conclusion 
that can be drawn from this comparison with {\em classical} bound OCs is that Cr\,197 
and vdB\,92 may be evolving into OB associations and/or towards dissolution. Given 
their young ages, the determining factor for this behaviour is certainly related to 
another mechanism than age-dependent dynamical evolution. Instead, it is probably 
associated with the primordial star formation process, or even earlier, to the molecular 
cloud fragmentation. Both clusters may have been formed with dynamically hot stellar
components, and thus, are unstable against dissolution (e.g. \citealt{GoodW09}). A 
similar conclusion was reached for some equally young objects, but with very 
different masses: NGC\,2244 (\citealt{N2244}) and Bochum\,1 (\citealt{Bochum1}) 
have comparable stellar masses ($\mcl\sim600\,\ms$), while NGC\,1931 and Pismis\,5 
(\citealt{Pi5}) are at the low-mass end ($\mcl\sim180\,\ms$, and $\mcl\sim60\,\ms$, 
respectively). Thus, early star cluster dissolution appears to leave detectable
imprints also on the structure (e.g. RDP) of clusters as massive as several hundreds 
solar masses.

\section{Summary and conclusions}
\label{Conclu}

Given the complex interplay among environmental conditions, star-formation efficiency, 
initial dynamical state of the stellar content, and the total mass converted into stars 
or expelled, the majority of the embedded clusters do not survive the first few tens of 
Myr, especially the low-mass ones. In this context, it is important to investigate the 
structural and photometric properties of OCs (with a mass range and in different 
environments) that are undergoing this early phase.

In the present paper we derive astrophysical parameters and investigate the nature 
of the young OCs Cr\,197 and vdB\,92. Their location in the $3^{rd}$ Galactic quadrant 
minimises contamination by field stars. We work with field-star decontaminated 2MASS 
photometry (with errors $\la0.1$\,mag), which enhances CMD evolutionary sequences and 
stellar RDPs, yielding more constrained fundamental and structural parameters. 

The decontaminated CMDs are characterised by similar properties, an under-populated
and developing MS, a dominant fraction ($\ga75\%$) of PMS stars, and some differential 
reddening. In both cases, the MS and PMS CMD morphologies consistently imply a time-spread 
of $\sim10$\,Myr in the star formation. Thus, we set the age of Cr\,197 and vdB\,92 as
constrained within $5\pm4$\,Myr. Their MS$+$PMS stellar masses are 
$\approx660^{+102}_{-59}\,\ms$ (Cr\,197) and $\approx750^{+101}_{-51}\,\ms$ (vdB\,92). By
means of the proper motion of the member MS and PMS stars, we estimate their velocity
dispersions to be in the range $\approx15\,\kms$ to $\approx22\,\kms$.

Compared to a set of {\em classical} bound OCs, both Cr\,197 and vdB\,92 appear to have 
core and cluster radii abnormally large, with $\rc\approx1.5,~1.9$\,pc and 
$\rl\approx12,~8$\,pc, respectively for Cr\,197 and vdB\,92. Structurally, the stellar 
RDPs follow a cluster-like profile for most of the radial range, except in the central
region, where they have a pronounced cusp. At less than about 10\,Myr, this cusp is probably 
related to the star formation and/or molecular cloud fragmentation, and not the product of 
dynamical evolution. A possible conclusion is that Cr\,197 and vdB\,92 deviate critically 
from dynamical equilibrium, and similarly to the equally young NGC\,2244 and Bochum\,1 (of 
comparable mass), and Pismis\,5 and NGC\,1931 (of significantly lower mass), they are heading 
towards dissolution. This interpretation is also consistent with the super-virial state of 
both clusters (as well as the dissolving OC NGC\,2244), which have velocity dispersions much 
higher than the $\sigma_v\approx1\,\kms$ expected for nearly virialised clusters of similar 
mass and size as Cr\,197 and vdB\,92. In this context, Cr\,197 and vdB\,92 may be taken as a 
link between embedded clusters and young stellar associations.

We provide evidence that early star cluster dissolution may be detected, for instance, 
by means of important and systematic irregularities in the stellar radial density profile 
of clusters as massive as several $10^2\,\ms$. Studies like the present one are important 
for a better understanding of the crucial early evolution of embedded star clusters - and 
the dependence on mass and environment - that rarely result in a {\em classical} bound OC 
or, more frequently, lead to their dissolution into the field. 

\section*{Acknowledgements}
We thank the referee, Simon Goodwin, for interesting comments and suggestions.
We acknowledge support from the Brazilian Institution CNPq.
This publication makes use of data products from the Two Micron All Sky Survey, which
is a joint project of the University of Massachusetts and the Infrared Processing and
Analysis Centre/California Institute of Technology, funded by the National Aeronautics
and Space Administration and the National Science Foundation. This research has made 
use of the WEBDA database, operated at the Institute for Astronomy of the University
of Vienna.

\label{lastpage}

\begin{thebibliography}{}

\bibitem[\protect\citeauthoryear{Allison et al.}{2009}]{Allison09}
   Allison R.J., Goodwin S.P., Parker R.J., de Grijs R., Portegies Zwart S.F.
   \& Kouwenhoven M.B.N. 2009, ApJL, 700,99
   
\bibitem[\protect\citeauthoryear{Alter et al.}{1970}]{Alter70}
   Alter G., Balazs B., Ruprecht J. \& Vanysek J. 1970, in {\em Catalogue of
   Star Clusters and Associations}, Budapest Akademiai Kiado, 2nd Edition, ed.
   by Alter G., Balazs  B. \& Ruprecht J.
   
\bibitem[\protect\citeauthoryear{Bastian et al.}{2005}]{Bast05}
   Bastian N., Gieles M., Lamers H.J.G.L.M., Scheepmaker R.A. \& de Grijs R. 
   2005, A\&A, 431, 905
   
\bibitem[\protect\citeauthoryear{Battinelli \& Capuzzo-Dolcetta}{1991}]{Bat91}
   Battinelli P. \& Capuzzo-Dolcetta R. 1991, MNRAS, 249, 76
   
\bibitem[\protect\citeauthoryear{Battinelli, Brandimarti \& Capuzzo-Dolcetta}{1994}]{Bat94}
   Battinelli P., Brandimarti A. \& Capuzzo-Dolcetta R. 1994, A\&AS, 104, 379
  
\bibitem[\protect\citeauthoryear{van den Bergh}{1966}]{vdB66}
   van den Bergh S. 1966, AJ, 71, 990
     
\bibitem[\protect\citeauthoryear{Bessel \& Brett}{1988}]{BesBret88}
   Bessel M.S. \& Brett J.M. 1988, PASP, 100, 1134
   
\bibitem[\protect\citeauthoryear{Bica, Bonatto \& Dutra}{2008}]{Bochum1}
   Bica E., Bonatto C. \& Dutra C. 2008, A\&A, 489, 1129

\bibitem[\protect\citeauthoryear{Bica et al.}{2006}]{GCProp}
   Bica E., Bonatto C., Barbuy B. \& Ortolani S. 2006, A\&A, 450, 105

\bibitem[\protect\citeauthoryear{Bica, Bonatto \& Camargo}{2008}]{ProbFSR}
   Bica E., Bonatto C. \& Camargo D. 2008, MNRAS, 385, 349

\bibitem[\protect\citeauthoryear{Bonatto, Bica \& Girardi}{2004}]{TheoretIsoc}
   Bonatto C., Bica E. \& Girardi L. 2004, A\&A, 415, 571

\bibitem[\protect\citeauthoryear{Bonatto \& Bica}{2005}]{DetAnalOCs}
   Bonatto C. \&  Bica E. 2005, A\&A, 437, 483

\bibitem[\protect\citeauthoryear{Bonatto, Santos Jr. \& Bica}{2006}]{N6611}
   Bonatto C., Santos Jr. J.F.C. \& Bica E. 2006, A\&A, 445, 567

\bibitem[\protect\citeauthoryear{Bonatto et al.}{2006}]{N4755}
   Bonatto C., Bica E., Ortolani S. \& Barbuy B. 2006, A\&A, 453, 121
   
\bibitem[\protect\citeauthoryear{Bonatto \& Bica}{2006}]{N3960}
   Bonatto C. \& Bica E. 2006, A\&A, 455, 931

\bibitem[\protect\citeauthoryear{Bonatto \& Bica}{2007}]{BB07}
   Bonatto C. \& Bica E. 2007, MNRAS, 377, 1301
   
\bibitem[\protect\citeauthoryear{Bonatto \& Bica}{2008}]{StrucPar}
   Bonatto C. \& Bica E. 2008, A\&A, 477, 829
   
\bibitem[\protect\citeauthoryear{Bonatto \& Bica}{2009a}]{LKstuff}
   Bonatto C. \& Bica E. 2009a, MNRAS, 392, 483
   
\bibitem[\protect\citeauthoryear{Bonatto \& Bica}{2009b}]{N2244}
   Bonatto C. \& Bica E. 2009b, MNRAS, 394, 2127
   
\bibitem[\protect\citeauthoryear{Bonatto \& Bica}{2009c}]{Pi5}
   Bonatto C. \& Bica E. 2009c, MNRAS, 397, 1915

\bibitem[\protect\citeauthoryear{Cardelli, Clayton \& Mathis}{1989}]{Cardelli89}
   Cardelli J.A., Clayton G.C. \& Mathis, J.S. 1989, ApJ, 345, 245
   
\bibitem[\protect\citeauthoryear{Chen, de Grijs \& Zhao}{2004}]{CGZ07}
   Chen L., de Grijs R. \& Zhao J.L. 2007, AJ, 134, 1368
   
\bibitem[\protect\citeauthoryear{Clari\'a}{1974a}]{Claria74a}
   Clari\'a J.J. 1974a, ApJ, 79, 1022

\bibitem[\protect\citeauthoryear{Clari\'a}{1974b}]{Claria74b}
   Clari\'a J.J. 1974b, A\&A, 37, 229
   
\bibitem[\protect\citeauthoryear{Collinder}{1931}]{Coll31}
   Collinder P. 1931, AnLun, 2, 1

\bibitem[\protect\citeauthoryear{Dutra, Santiago \& Bica}{2002}]{DSB2002}
   Dutra C.M., Santiago B.X. \& Bica E. 2002, A\&A, 383, 219
      
\bibitem[\protect\citeauthoryear{Friel}{1995}]{Friel95}
   Friel E.D. 1995, ARA\&A 1995, 33, 381
   
\bibitem[\protect\citeauthoryear{Froebrich, Scholz \& Raftery}{2007}]{Froeb07}
   Froebrich D., Scholz A. \& Raftery C.L. 2007, MNRAS, 374, 399
   
\bibitem[\protect\citeauthoryear{Furlan et al.}{2009}]{Furlan09}
   Furlan E., Watson D.M., McClure M.K., Manoj, P., Espaillat C., D'Alessio P.,
   Calvet N., Kim K.H. et al. 2009, ApJ, 703, 1964
   
\bibitem[\protect\citeauthoryear{Ghez et al.}{2008}]{Ghez08}
   Ghez A.M., Salim S., Weinberg N.N., Lu J.R., Do T., Dunn J.K., Matthews K.,
   Morris M.R. et al. 2008, ApJ, 689, 1044
   
\bibitem[\protect\citeauthoryear{Gieles, Sana \& Portegies Zwart}{2008}]{GSPZ10}
   Gieles M., Sana H. \& Portegies Zwart S.F. 2010, MNRAS, 402, 1750
   
\bibitem[\protect\citeauthoryear{Girardi et al.}{2002}]{Girardi2002}
   Girardi L., Bertelli G., Bressan A., Chiosi C., Groenewegen M.A.T.,
   Marigo P., Salasnich B. \& Weiss A. 2002, A\&A, 391, 195  
     
\bibitem[\protect\citeauthoryear{Goodwin}{2009}]{GoodW09}
   Goodwin S.P. 2009, Ap\&SS, 324, 259
   
\bibitem[\protect\citeauthoryear{Goodwin \& Bastian}{2006}]{GoBa06}
   Goodwin S.P. \& Bastian N. 2006, MNRAS, 373, 752
   
\bibitem[\protect\citeauthoryear{Gouliermis et al.}{2000}]{Goul00}
   Gouliermis D., Kontizas M., Korakitis R., Morgan D.H., Kontizas E.
   \& Dapergolas A. 2000, AJ, 119, 1737
   
\bibitem[\protect\citeauthoryear{de Grijs, Kouwenhoven \& Goodwin}{2008}]{GKG08}
   de Grijs R., Kouwenhoven M.B.N. \& Goodwin S.P. 2008, AN, 329, 972
   
\bibitem[\protect\citeauthoryear{de Grijs \& Goodwin}{2008}]{deGG08}
   de Grijs R. \& Goodwin S.P. 2008, MNRAS, 383, 1000
   
\bibitem[\protect\citeauthoryear{de Grijs \& Goodwin}{2009}]{deGG09}
   de Grijs R. \& Goodwin S.P. 2009, in {\em IAU Symposium 256}, van 
   Loon J.T. \& Oliveira J.M., eds., Cambridge Univ. Press
   
\bibitem[\protect\citeauthoryear{Gum}{1955}]{Gum55}
   Gum C.S. 1955, MmRAS, 67, 155
   
\bibitem[\protect\citeauthoryear{King}{1962}]{King1962}
   King I. 1962, AJ, 67, 471
   
\bibitem[\protect\citeauthoryear{Kroupa}{2001}]{Kroupa2001}
   Kroupa P. 2001, MNRAS, 322, 231
      
\bibitem[\protect\citeauthoryear{Lada \& Lada}{2003}]{LL2003}
   Lada C.J. \& Lada E.A. 2003, ARA\&A, 41, 57
   
\bibitem[\protect\citeauthoryear{Lauberts}{1982}]{Lauberts82}
   Lauberts A. 1982, ESO/Uppsala survey of the ESO(B) atlas, European Southern
   Observatory
   
\bibitem[\protect\citeauthoryear{Magakian}{2003}]{Magakian03}
   Magakian T.Y. 2003, A\&A, 399, 141
   
\bibitem[\protect\citeauthoryear{Massey, Johnson \& Gioia-Eastwood}{1995}]{Massey95}
   Massey P., Johnson K.E. \& De Gioia-Eastwood K. 1995, ApJ, 454, 151

\bibitem[\protect\citeauthoryear{Momany et al.}{2006}]{Momany06}
   Momany Y., Zaggia S., Gilmore G., Piotto G., Carraro G., Bedin
   L.R. \& de Angeli F. 2006, A\&A, 451, 515
   
\bibitem[\protect\citeauthoryear{Naylor \& Jeffries}{2006}]{NJ06}
   Naylor T. \& Jeffries R.D. 2006, MNRAS, 373, 1251
   
\bibitem[\protect\citeauthoryear{Pettersson \& Reipurth}{1994}]{PetBo94}
   Pettersson B. \& Reipurth B. 1994, A\&ASS, 104, 233
   
\bibitem[\protect\citeauthoryear{Piskunov et al.}{2007}]{Piskunov07}
   Piskunov A.E., Schilbach E., Kharchenko N.V., R\"oser S. \& Scholz R.-D.
   2007, A\&A, 468, 151
   
\bibitem[\protect\citeauthoryear{Poetzel, Mundt \& Ray}{1989}]{Poetzel89}
   Poetzel R., Mundt R. \& Ray T.P. 1989, A\&A, 224, L13
   
\bibitem[\protect\citeauthoryear{Rodgers, Campbell \& Whiteoak}{1960}]{RCW60}
   Rodgers A.W., Campbell C.T. \& Whiteoak, J.B. 1960, MNRAS, 121, 103
   
\bibitem[\protect\citeauthoryear{Siess, Dufour \& Forestini}{2000}]{Siess2000}
   Siess L., Dufour E. \& Forestini M. 2000, A\&A, 358, 593

\bibitem[\protect\citeauthoryear{Skrutskie et al.}{1997}]{2mass1997}
   Skrutskie M., Schneider S.E., Stiening R., Strom S.E., Weinberg M.D.,
   Beichman C., Chester T., Cutri R .et al. 1997, in {\it The Impact
   of Large Scale Near-IR Sky Surveys}, ed. F. Garzon et al., Kluwer 
   (Netherlands), 210, 187
   
\bibitem[\protect\citeauthoryear{Soares \& Bica}{2003}]{Soares03}
   Soares J.B. \& Bica E. 2003, A\&A, 404, 217
   
\bibitem[\protect\citeauthoryear{Spitzer}{1987}]{Spitzer87}
   Spitzer L. 1987, in {\em Dynamical Evolution of Globular Clusters},
   Princeton, NJ, Princeton University Press, p. 191
   
\bibitem[\protect\citeauthoryear{Trager, King \& Djorgovski}{1995}]{TKD95}
   Trager S.C., King I.R. \& Djorgovski S. 1995, AJ, 109, 218
   
\bibitem[\protect\citeauthoryear{Trippe et al.}{2008}]{Trippe08}
   Trippe S., Gillessen S., Gerhard O.E., Bartko H., Fritz T.K., Maness H.L.
   Eisenhauer F., Martins F., et al.  2008, A\&A, 492, 419
   
\bibitem[\protect\citeauthoryear{Tutukov}{1978}]{tutu78}
   Tutukov A.V. 1978, A\&A, 70, 57
   
\bibitem[\protect\citeauthoryear{Vogt \& Moffat}{1973}]{VM73}
   Vogt N. \& Moffat A.F.J. 1973, A\&AS, 9, 97
   
\bibitem[\protect\citeauthoryear{}{2007}]{Whit07}
   Whitmore B.C., Chandar R. \& Fall S.M. 2007, AJ, 133, 1067
   
\bibitem[\protect\citeauthoryear{Zacharias et al.}{2009}]{Zach09}
   Zacharias N., Finch C., Girard T., Hambly N., Wycoff G.,
   Zacharias M.I. , Castillo D., Corbin T. et al. 2009, AJ, submitted
   
\end{thebibliography}
\end{document}